\documentclass[twocolumn,aps,showpacs,superscriptaddress,tightenlines,a4paper]{revtex4}

\usepackage{amssymb}
\usepackage{amsmath}
\usepackage{bbold}
\usepackage{mathrsfs}
\usepackage{paralist}
\usepackage{graphicx}
\usepackage{xcolor}

\begin{document}
%
%
	\title{
	Canonical quantization of the electromagnetic field interacting with a moving dielectric
	}
%
%
	\author{S. A. R. Horsley}
	\affiliation{School of Physics and Astronomy, University of St Andrews,
North Haugh, St Andrews, KY16 9SS, UK.}
	\email{sarh@st-andrews.ac.uk}
%
%
	\begin{abstract}
		The electromagnetic field is canonically quantized in the presence of a linear, dispersive and dissipative medium that is in uniform motion.  Specifically we calculate the change in the normal modes of the coupled matter--field system and find a Hamiltonian that contains negative energy normal modes.  We interpret these modes as the origin of phenomena such as quantum friction, and find that a detector initially in its ground state and coupled to the electromagnetic field in the vicinity of, or within a uniformly moving medium has a non--zero probability of excitation at \(T=0\,\text{K}\).
	\end{abstract}
%
%
%
	\pacs{03.70.+k,12.20-m,42.50.-p,03.50.De,03.30.+p}
	\maketitle
	\bibliographystyle{unsrt}
	\par
%
%
	\section{Introduction}
	\par
	In quantum field theory, time and space play quite different roles.  One consequence of this is the production of particles in the presence of moving boundaries, which is particularly evident in the predicted response of the vacuum electromagnetic field to a dielectric medium in motion: nonuniformly accelerated mirrors should radiate~\cite{fulling__1976}, and a process analogous to that of Hawking emission~\cite{hawking__1975} has been predicted to occur when a localized change in the refractive index moves through a medium at a speed exceeding the phase velocity of light in the medium~\cite{schutzhold__2005,philbin__2008,belgiorno__2010}.  Indeed, it is surprising is that there should be effects of this type that occur for dielectrics in \emph{uniform} motion.  Nevertheless, one such effect is that of quantum friction, whereby two dielectric plates in relative lateral motion are predicted to come to rest due a production of pairs of surface excitations out of the vacuum state~\cite{pendry__1997}.
	\par
	The physics of these phenomena been the subject of recent controversy, with some debate concerning the reality of quantum friction~\cite{pendry__2010,leonhardt__2010,barton2010a,barton2010b,volokitin2011}, as well as difficulty in quantitatively interpreting the experimental observation of the aforementioned Hawking like process~\cite{belgiorno__2010,schutzhold__2011,belgiorno__2011}.  It seems like part of this discord arises from a lack of agreement about the theory that should be used to describe quantum electromagnetism interacting with a moving dielectric.  Aspects of the quantization of electromagnetism in moving media have been considered before, initially by Jauch and Watson~\cite{jauch__1948}, but as far as the author is aware, there has been no canonical approach that includes the effects of dispersion and dissipation~\footnote{An attempt to formulate such a theory was made in~\cite{amooshahi__2009}, however this theory does not appear to correctly account for the Doppler effect.}.  Here we develop a canonical theory that includes these effects and we apply it to some simple physical scenarios.
	\par
	The point of view taken here is that dispersion and dissipation are not merely complications that somehow superimpose onto a `clean' case where a dielectric can be represented by a refractive index, \(n\) that is independent of frequency, but are in fact fundamental to the interaction between macroscopic objects and the quantized electromagnetic field.  After all, the fluctuation--dissipation theorem tells us that for linear dielectrics we can always understand the vacuum electromagnetic field as being produced by currents within the dielectric that are in proportion to the square root of the dissipative response times \(\hbar\)~\cite{volume5}.  To develop a canonical theory of electromagnetism that is consistent with this perspective, we use a method similar to that of Huttner and Barnett~\cite{huttner__1992}, where a fictional reservoir is added to the electromagnetic Hamiltonian to account for both the absorbed field energy and the dispersive response.  Philbin has recently presented a modification and generalization of this procedure that applies to media that can be described by any \(\epsilon(\boldsymbol{x},\omega)\) and \(\mu(\boldsymbol{x},\omega)\) that satisfy the Kramers--Kronig relations~\cite{philbin__2010}.  Philbin's approach was extended within~\cite{horsley__2011-1}, where the Lagrangian necessary to describe moving media was derived, the quantization of which we now investigate.\\
%
%
	\section{Classical electromagnetism interacting with a moving medium\label{sec1}}
	\par
	To begin, we give the form of the action necessary to derive the classical equations for the electromagnetic field interacting with a moving medium, as presented in~\cite{horsley__2011-1}.  This is given as the integral of a Lagrangian density over space--time, \(S[A^{\mu},\boldsymbol{X}_{\omega},\boldsymbol{Y}_{\omega}]=\int d^{4}x\mathscr{L}\), where
	\begin{equation}
		\mathscr{L}=\mathscr{L}_{\text{\tiny{F}}}+\mathscr{L}_{\text{\tiny{R}}}+\mathscr{L}_{\text{\tiny{INT}}}\label{lagrangian-density}
	\end{equation}
	The part of the Lagrangian associated with the degrees of freedom within the electromagnetic field is of the usual Lorentz--invariant form
	\begin{equation}
		\mathscr{L}_{\text{\tiny{F}}}=\frac{\epsilon_{0}}{2}\left[\boldsymbol{E}^{2}-c^{2}\boldsymbol{B}^{2}\right]
	\end{equation}
	where the fields are written in terms of the potentials \((\varphi,\boldsymbol{A})\) as \(\boldsymbol{E}=-\boldsymbol{\nabla}\varphi-\dot{\boldsymbol{A}}\) and \(\boldsymbol{B}=\boldsymbol{\nabla}\boldsymbol{\times}\boldsymbol{A}\).  The electromagnetic field is coupled to two continua of oscillators, \(\boldsymbol{X}_{\omega}\) \& \(\boldsymbol{Y}_{\omega}\) that represent the collective degrees of freedom of the medium.  The coupling of the field and the oscillators is chosen so as to reproduce the experimentally measured linear susceptibilities of the dielectric from the equations of motion.  For a dielectric moving with velocity \(\boldsymbol{V}\), the part of the Lagrangian describing the oscillators is modified from that of~\cite{philbin__2010} due to the Lorentz transformation of the time coordinate
	\begin{multline}
		\mathscr{L}_{\text{\tiny{R}}}=\frac{1}{2}\int_{0}^{\infty}d\omega\left\{\gamma^{2}\left[\dot{\boldsymbol{X}}_{\omega}+\left(\boldsymbol{V}\boldsymbol{\cdot}\boldsymbol{\nabla}\right)\boldsymbol{X}_{\omega}\right]^{2}-\omega^{2}\boldsymbol{X}_{\omega}^{2}\right\}\\
		+(\boldsymbol{X}_{\omega}\to\boldsymbol{Y}_{\omega})\label{lag-res}
	\end{multline}
	where \(\gamma=(1-\boldsymbol{V}^{2}/c^{2})^{-1/2}\).  Additionally, the transformation of the field amplitudes means that the coupling between matter and electromagnetism involves tensors, \(\boldsymbol{\alpha}_{\text{\tiny{EB}}}\) \& \(\boldsymbol{\alpha}_{\text{\tiny{BE}}}\), in addition to those of~\cite{philbin__2010}
	\begin{multline}
	\mathscr{L}_{\text{\tiny{INT}}}=\boldsymbol{E}\boldsymbol{\cdot}\int_{0}^{\infty}d\omega\left[\boldsymbol{\alpha}_{\text{\tiny{EE}}}\cdot\boldsymbol{X}_{\omega}+\boldsymbol{\alpha}_{\text{\tiny{EB}}}\cdot\boldsymbol{Y}_{\omega}\right]\\
	+\boldsymbol{B}\boldsymbol{\cdot}\int_{0}^{\infty}d\omega\left[\boldsymbol{\alpha}_{\text{\tiny{BB}}}\cdot\boldsymbol{Y}_{\omega}+\boldsymbol{\alpha}_{\text{\tiny{BE}}}\cdot\boldsymbol{X}_{\omega}\right].\label{lint}
	\end{multline}
	where the \(\boldsymbol{\alpha}\) tensors are typically functions of \(\omega\), \(\boldsymbol{x}\), and \(t\), even if they are only functions of frequency and position in the rest frame.
	\par
	Considering the specific case where the medium is described in terms of a scalar permittivity and permeability in the rest frame, and the motion is \(\boldsymbol{V}=V\hat{\boldsymbol{x}}\), the coupling tensors take the form; \(\boldsymbol{\alpha}_{\text{\tiny{EE}}}(\boldsymbol{x},\omega,t)=\boldsymbol{\Lambda}\alpha(\boldsymbol{x}^{\prime},\omega)\); \(\boldsymbol{\alpha}_{\text{\tiny{BB}}}(\boldsymbol{x},\omega,t)=\boldsymbol{\Lambda}\beta(\boldsymbol{x}^{\prime},\omega)\); \(\boldsymbol{\alpha}_{\text{\tiny{EB}}}(\boldsymbol{x},\omega,t)=\gamma\beta(\boldsymbol{x}^{\prime},\omega)\boldsymbol{\mathbb{1}}_{3}\boldsymbol{\times}\boldsymbol{V}/c^{2}\); and \(\boldsymbol{\alpha}_{\text{\tiny{BE}}}(\boldsymbol{x},\omega,t)=-\gamma\alpha(\boldsymbol{x}^{\prime},\omega)\boldsymbol{\mathbb{1}}_{3}\boldsymbol{\times}\boldsymbol{V}\), where \(\alpha(\boldsymbol{x}^{\prime},\omega)=\sqrt{2\omega\text{Im}[\chi_{\text{\tiny{EE}}}(\boldsymbol{x}^{\prime},\omega)]/\pi}\), \(\beta(\boldsymbol{x}^{\prime},\omega)=\sqrt{2\omega\text{Im}[\chi_{\text{\tiny{BB}}}(\boldsymbol{x}^{\prime},\omega)]/\pi}\), and \(\boldsymbol{\Lambda}=\text{diag}(1,\gamma,\gamma)\).  The susceptibilities, \(\chi_{\text{\tiny{EE}}}\) \& \(\chi_{\text{\tiny{BB}}}\) are those of the medium in the rest frame, and \(\boldsymbol{x}^{\prime}\) is the rest frame spatial coordinate: \(x^{\prime}=\gamma(x-Vt)\); \(y^{\prime}=y\); and \(z^{\prime}=z\).  As is well known (e.g.~\cite{matloob2005b}), this modification of the coupling tensors turns an isotropic medium into an anisotropic one when it is in motion.  Furthermore, if the medium is inhomogeneous in the rest frame, then it will be time dependent in the laboratory frame.  Only when the motion is along an axis where the medium has translational symmetry will this not be the case.
%
%
	\subsection{Solutions to the equations of motion\label{clas-sol-sec}}
	\par
	The solutions to the classical equations of motion of (\ref{lagrangian-density}) form the basis of the quantization procedure, and we will expand the field operators in terms of these functions in the next section.
	\par
	The equations of motion for the continua of oscillators can be found from the usual formulae,
	\begin{equation}
		\frac{\partial}{\partial x^{\mu}}\left(\frac{\partial\mathscr{L}}{\partial_{\mu}\boldsymbol{X}_{\omega}}\right)=\frac{\partial\mathscr{L}}{\partial\boldsymbol{X}_{\omega}}
	\end{equation}
	which gives
	\begin{equation}
		\left[\gamma^{2}\left(\frac{\partial}{\partial t}+\boldsymbol{V}\boldsymbol{\cdot}\boldsymbol{\nabla}\right)^{2}+\omega^{2}\right]\boldsymbol{X}_{\omega}=\boldsymbol{\alpha}_{\text{\tiny{EE}}}^{T}\boldsymbol{\cdot}\boldsymbol{E}+\boldsymbol{\alpha}_{\text{\tiny{BE}}}^{T}\boldsymbol{\cdot}\boldsymbol{B}\label{osc1}
	\end{equation}
	with the \(\boldsymbol{Y}_{\omega}\) obeying the same equation after the substitutions  \(E\leftrightarrow B\) and \(X\leftrightarrow Y\).  Throughout the paper we shall most often only explicitly write out the quantities associated with the \(\boldsymbol{X}_{\omega}\) and the polarization, \(\boldsymbol{P}\), but always with the understanding that the remaining results can be obtained using this substitution.
	\par
	The equations of motion for the electromagnetic field are found to be the usual macroscopic equations
	\begin{align}
	\boldsymbol{\nabla}\boldsymbol{\cdot}\left(\epsilon_{0}\boldsymbol{E}+\boldsymbol{P}\right)&=0\nonumber\\
	\boldsymbol{\nabla}\boldsymbol{\times}\boldsymbol{B}-\frac{1}{c^{2}}\frac{\partial\boldsymbol{E}}{\partial t}&=\mu_{0}\left(\frac{\partial\boldsymbol{P}}{\partial t}+\boldsymbol{\nabla}\boldsymbol{\times}\boldsymbol{M}\right)\label{field-eqn}
	\end{align}
	where \(\boldsymbol{P}=\int_{0}^{\infty}d\omega\left[\boldsymbol{\alpha}_{\text{\tiny{EE}}}\cdot\boldsymbol{X}_{\omega}+\boldsymbol{\alpha}_{\text{\tiny{EB}}}\cdot\boldsymbol{Y}_{\omega}\right]\), and  \(\boldsymbol{M}\) takes a similar form, but with the aforementioned substitutions.
	\par
	\begin{figure}[h!]
		\includegraphics[width=6cm]{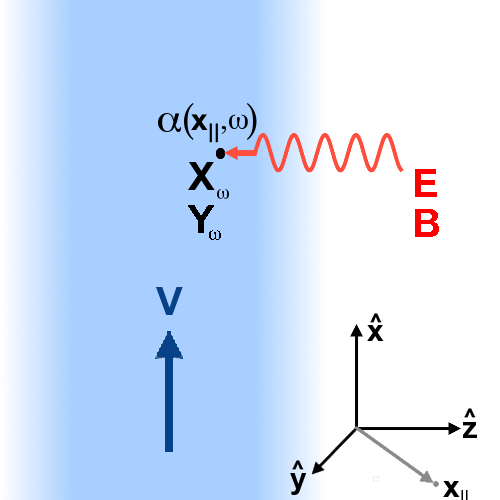}
		\caption{The electromagnetic field \((\boldsymbol{E},\boldsymbol{B})\) is coupled to two continua of oscillators \((\boldsymbol{X}_{\omega},\boldsymbol{Y}_{\omega})\).  The strength of the coupling is given in terms of the set of \(\boldsymbol{\alpha}\) tensors within (\ref{lint}), and the interaction of the field and these oscillators is such that the correct constitutive relations emerge from the equations of motion.  In our case we are investigating a dielectric in motion along \(\hat{\boldsymbol{x}}\), that is allowed to be inhomogeneous only as a function of \(\boldsymbol{x}_{\parallel}\).\label{fig1}}
	\end{figure}
	\par
	With the simplification of translational symmetry along the direction of motion, \(\hat{\boldsymbol{x}}\) (see figure \ref{fig1}), the solution of (\ref{osc1}) becomes
	\begin{multline}
		\tilde{\boldsymbol{X}}_{\omega}(k,\boldsymbol{x}_\parallel,\Omega)=\frac{\boldsymbol{\alpha}_{\text{\tiny{EE}}}^{T}(\boldsymbol{x}_{\parallel},\omega)\boldsymbol{\cdot}\tilde{\boldsymbol{E}}+\boldsymbol{\alpha}_{\text{\tiny{BE}}}^{T}(\boldsymbol{x}_{\parallel},\omega)\boldsymbol{\cdot}\tilde{\boldsymbol{B}}}{(\omega-\Omega_{-}-i\eta)(\omega+\Omega_{-}+i\eta)}\\
		+\tilde{\boldsymbol{X}}_{0\omega}(k,\boldsymbol{x}_\parallel,\Omega)\label{simple-osc-sol}
	\end{multline}
	the tilde indicating a Fourier transform with respect to \(x\), and \(t\).  In (\ref{simple-osc-sol}) we have introduced the following  new quantities: \(\tilde{\boldsymbol{X}}_{0\omega}=2\pi[\delta(\Omega_{-}-\omega)\boldsymbol{h}_{X_{\omega}}(k,\boldsymbol{x}_{\parallel})+\delta(\Omega_{-}+\omega)\boldsymbol{h}^{\star}_{X_{\omega}}(-k,\boldsymbol{x}_{\parallel})]\), \(\Omega_{\pm}=\gamma(\Omega\pm\boldsymbol{V}\boldsymbol{\cdot}\boldsymbol{k})\), the vector \(\boldsymbol{x}_{\parallel}\) is the coordinate in the \(y-z\) plane, and \(\eta\) is an infinitesimal quantity that specifies the \emph{retarded} solution of the driven oscillator equation (\ref{osc1}).  The expression given by (\ref{simple-osc-sol}) should be understood in the limit \(\eta\to0\) and be evaluated in terms of poles and principal parts.  The (classical) amplitudes, \(\boldsymbol{h}_{X_{\omega}}\) may be freely specified, and represent the motion of the oscillators that is not driven by the electromagnetic field.
	\par
	Inserting (\ref{simple-osc-sol}) into (\ref{field-eqn}) we identify the electromagnetic susceptibilities as follows
	\begin{align}
	\boldsymbol{\chi}_{\text{\tiny{EE}}}(\Omega_{-})&=\int_{0}^{\infty}\frac{\boldsymbol{\alpha}_{\text{\tiny{EE}}}(\omega)\boldsymbol{\cdot}\boldsymbol{\alpha}_{\text{\tiny{EE}}}^{T}(\omega)+\boldsymbol{\alpha}_{\text{\tiny{EB}}}(\omega)\boldsymbol{\cdot}\boldsymbol{\alpha}_{\text{\tiny{EB}}}^{T}(\omega)}{(\omega-\Omega_{-}-i\eta)(\omega+\Omega_{-}+i\eta)}d\omega\nonumber\\
	\boldsymbol{\chi}_{\text{\tiny{EB}}}(\Omega_{-})&=\int_{0}^{\infty}\frac{\boldsymbol{\alpha}_{\text{\tiny{EE}}}(\omega)\boldsymbol{\cdot}\boldsymbol{\alpha}_{\text{\tiny{BE}}}^{T}(\omega)+\boldsymbol{\alpha}_{\text{\tiny{EB}}}(\omega)\boldsymbol{\cdot}\boldsymbol{\alpha}_{\text{\tiny{BB}}}^{T}(\omega)}{(\omega-\Omega_{-}-i\eta)(\omega+\Omega_{-}+i\eta)}d\omega\label{susdef}
	\end{align}
	Notice that due to the additional terms depending on \(\boldsymbol{V}\) within (\ref{lag-res}), the Doppler shifted frequency, \(\Omega_{-}\) appears within the susceptibilities.  Furthermore, once (\ref{susdef}) has been expanded in terms of poles and principal parts, it is found that the real and imaginary parts of these susceptibilities are related by the Kramers--Kronig relations~\cite{horsley__2011-1}.   Using (\ref{simple-osc-sol}) and the counterpart expression for the \(\boldsymbol{Y}_{\omega}\), the \(\boldsymbol{X}_{\omega}\) \& \(\boldsymbol{Y}_{\omega}\) can be eliminated from (\ref{field-eqn}).  The Maxwell equations then become those of macroscopic electromagnetism in a moving dielectric, defined in terms of the susceptibilities given in (\ref{susdef}), and with `free' current and charge densities proportional to \(\boldsymbol{h}_{X_{\omega}}\) \& \(\boldsymbol{h}_{Y_{\omega}}\).  The formal solution of these equations may be obtained through finding an appropriate Green function.  We thereby find the following expression for the electric field
	\begin{equation}
		\tilde{\boldsymbol{E}}(k,\boldsymbol{x}_{\parallel},\Omega)=i\Omega\int d^{2}\boldsymbol{x}_{\parallel}^{\prime}\boldsymbol{G}(k,\boldsymbol{x}_{\parallel},\boldsymbol{x}_{\parallel}^{\prime},\Omega)\boldsymbol{\cdot}\tilde{\boldsymbol{j}}_{0}(k,\boldsymbol{x}_{\parallel}^{\prime},\Omega)\label{Esol}
	\end{equation}
	where the Green function, \(\boldsymbol{G}(k,\boldsymbol{x}_{\parallel},\boldsymbol{x}_{\parallel}^{\prime},\Omega)\) is the solution of (\ref{field-G}).  The quantity appearing as a current in (\ref{Esol}) is given by
	\[
		\tilde{\boldsymbol{j}}_{0}(k,\boldsymbol{x}_{\parallel},\Omega)=-i\Omega\tilde{\boldsymbol{P}}_{0}(k,\boldsymbol{x}_{\parallel},\Omega)+\boldsymbol{\nabla}\boldsymbol{\times}\tilde{\boldsymbol{M}}_{0}(k,\boldsymbol{x}_{\parallel},\Omega)
	\]
	where we have defined, \(\tilde{\boldsymbol{P}}_{0}=\int_{0}^{\infty}d\omega[\boldsymbol{\alpha}_{\text{\tiny{EE}}}\boldsymbol{\cdot}\tilde{\boldsymbol{X}}_{0\omega}+\boldsymbol{\alpha}_{\text{\tiny{EB}}}\boldsymbol{\cdot}\tilde{\boldsymbol{Y}}_{0\omega}]\), and \(\boldsymbol{\nabla}=k\hat{\boldsymbol{x}}+{\boldsymbol{\nabla}}_{\parallel}\).  To complete the specification of the dynamics of the system, we can use the Maxwell equation, \(\boldsymbol{\nabla}\boldsymbol{\times}\tilde{\boldsymbol{E}}=i\Omega\tilde{\boldsymbol{B}}\), and (\ref{Esol}) to determine the magnetic field.\\
%
%
	\subsection{Classical Hamiltonian\label{classical-hamiltonian}}
	\par
	As the Lagrangian given in (\ref{lagrangian-density}) is local and contains only first order time derivatives, there therefore exists an associated Hamiltonian which we now derive.
	\par
	For the electromagnetic field the vector potential is the only dynamical variable and the associated canonical momentum density is given by
	\[
		\boldsymbol{\Pi}_{\boldsymbol{A}}=\frac{\partial\mathscr{L}}{\partial\dot{\boldsymbol{A}}}=-\epsilon_{0}\boldsymbol{E}-\boldsymbol{P}.
	\]
	while the oscillators have the associated momenta
	\begin{equation}
		\boldsymbol{\Pi}_{\boldsymbol{X}_{\omega}}=\frac{\partial\mathscr{L}}{\partial\dot{\boldsymbol{X}}_{\omega}}=\gamma^{2}[\dot{\boldsymbol{X}}_{\omega}+\left(\boldsymbol{V}\cdot\boldsymbol{\nabla}\right)\boldsymbol{X}_{\omega}].\label{Xcanmom}
	\end{equation}
	Due to the fact that electromagnetism has a gauge symmetry, the scalar potential has a canonical momentum equal to zero, \(\Pi_{\varphi}=\partial\mathscr{L}/\partial\dot{\varphi}=0\), and is not a dynamical field.  We must therefore interpret the first of (\ref{field-eqn})---which is not a dynamical equation---as a constraint that determines \(\varphi\) after some choice of gauge has been made.
	\par
	From these quantities, we can form the Hamiltonian density
	\begin{equation}
		\mathscr{H}=\boldsymbol{\Pi}_{\boldsymbol{A}}\boldsymbol{\cdot}\dot{\boldsymbol{A}}+\int_{0}^{\infty}d\omega\left[\boldsymbol{\Pi}_{\boldsymbol{X}_{\omega}}\boldsymbol{\cdot}\dot{\boldsymbol{X}}_{\omega}+\boldsymbol{\Pi}_{\boldsymbol{Y}_{\omega}}\boldsymbol{\cdot}\dot{\boldsymbol{Y}}_{\omega}\right]-\mathscr{L}
	\end{equation}
	which we give as the sum of two parts
	\begin{equation}
		\mathscr{H}=\mathscr{H}_{\text{\tiny{F}}}+\mathscr{H}_{\text{\tiny{R}}}\label{hamiltonian}
	\end{equation}
	the first of these is
	\begin{equation}
		\mathscr{H}_{\text{\tiny{F}}}=\frac{1}{2\epsilon_{0}}\left(\boldsymbol{\Pi}_{\boldsymbol{A}}+\boldsymbol{P}\right)^{2}
		-\left(\boldsymbol{\nabla}\boldsymbol{\times}\boldsymbol{A}\right)\boldsymbol{\cdot}\boldsymbol{M}+\frac{1}{2\mu_{0}}\left(\boldsymbol{\nabla}\boldsymbol{\times}\boldsymbol{A}\right)^{2}\label{field-hamiltonian}
	\end{equation}
	and the second is
	\begin{multline}
		\mathscr{H}_{\text{\tiny{R}}}=\frac{1}{2}\int_{0}^{\infty}\bigg[\frac{\boldsymbol{\Pi}_{\boldsymbol{X}_{\omega}}^{2}}{\gamma^{2}}+\omega^{2}\boldsymbol{X}_{\omega}^{2}-2\boldsymbol{V}\boldsymbol{\cdot}\left(\boldsymbol{\nabla}\boldsymbol{\otimes}\boldsymbol{X}_{\omega}\right)\boldsymbol{\cdot}\boldsymbol{\Pi}_{\boldsymbol{X}_{\omega}}\bigg]d\omega\label{oscillator-hamiltonian}\\
		+(\boldsymbol{X}_{\omega}\to\boldsymbol{Y}_{\omega}).
	\end{multline}
	Expression (\ref{field-hamiltonian}) was obtained through neglecting a term equal to a divergence (that will integrate to zero in the Hamiltonian) and then setting \(\varphi\boldsymbol{\nabla}\boldsymbol{\cdot}\boldsymbol{\Pi}_{\boldsymbol{A}}=0\).  This is the aforementioned constraint on \(\varphi\): \(\boldsymbol{\nabla}\boldsymbol{\cdot}\left[\epsilon_{0}\boldsymbol{E}+\boldsymbol{P}\right]=0\).  Finally the Hamiltonian, \(H=H_{\text{\tiny{F}}}+H_{\text{\tiny{R}}}\) is obtained from the integral of (\ref{hamiltonian}) over all space.  Notice that when \(\boldsymbol{V}\) is set to zero we obtain the Hamiltonian of~\cite{philbin__2010}.
\par	
	 The effect of the motion of the medium in (\ref{field-hamiltonian}--\ref{oscillator-hamiltonian}) is twofold.  Firstly, within (\ref{field-hamiltonian}) the canonical field momentum, \(\boldsymbol{\Pi}_{\boldsymbol{A}}\) is coupled to the \(\boldsymbol{Y}_{\omega}\) by the velocity, and \(\boldsymbol{A}\) is similarly coupled to \(\boldsymbol{X}_{\omega}\).  This is due to the Lorentz transformation of the field amplitudes, and means that the medium appears as a magnetoelectric (i.e. the electric polarization responds to the magnetic field as well as the electric field).
	 \par
	 Secondly, and more relevant to this discussion, the oscillator amplitudes within (\ref{oscillator-hamiltonian}) are coupled to the associated canonical momenta by the motion, via the term \(\Delta H=-\int_{0}^{\infty}\boldsymbol{V}\boldsymbol{\cdot}\left(\boldsymbol{\nabla}\boldsymbol{\otimes}\boldsymbol{X}_{\omega}\right)\boldsymbol{\cdot}\boldsymbol{\Pi}_{\boldsymbol{X}_{\omega}}d\omega+(\boldsymbol{X}_{\omega}\to\boldsymbol{Y}_{\omega})\).  This contribution has the physical meaning of the moving medium responding to the Doppler shifted frequencies of the electromagnetic field.  Indeed, if we want the susceptibilities (\ref{susdef}) to contain \(\Omega_{-}\) rather than \(\Omega\) then we have no choice but to include \(\Delta H\) in the expression for the energy of the system~\cite{horsley__2011-1}.  Yet this term is peculiar, and causes the Hamiltonian to lack a lower bound: the spatial dependence of \(\boldsymbol{X}_{\omega}\) can be made arbitrarily sharp so as to reduce \(\Delta H\) to an arbitrarily negative value with no change in the rest of the Hamiltonian.  Therefore, in quantum theory we should expect that the zero particle state will not be the lowest energy state of the system.  Indeed, when we come to describe the system with quantum mechanics, we will find that even at \(T=0 \text{K}\), it is possible to extract energy from the vacuum fluctuations of the electromagnetic field outside of a uniformly moving dielectric.  In appendix \ref{appendix-C} it is demonstrated that \(\Delta H\) is an accounting device that arises from enforcing a non--zero and uniform velocity that disappears once we include the centre of mass of the dielectric as a dynamical variable.  Throughout this work we take the simplest case where we suppose that the velocity of the dielectric has been fixed to some constant value by an external force.
	 \par
	 As a final comment, we note that in reality the energy of a moving dielectric interacting with the electromagnetic field cannot be reduced to \(-\infty\).  Spatial dispersion would become relevant at some large enough magnitude of \(\Delta H\), and provide a lower bound.  Mathematically this would be evident through a dependence on \(\boldsymbol{\nabla}\boldsymbol{\otimes}\boldsymbol{X}_{\omega}\) in the rest of the Hamiltonian.  This is an important approximation that does not seem to be mentioned within the existing literature.  However, although this warrants further investigation, here it does not alter our conclusions so long as we are careful, it just means that in reality for a large enough velocity the Hamiltonian can be reduced to a very large negative value rather than \(-\infty\).  We can avoid this complication for the time being so long as we ultimately avoid expressions that depend on wavelengths of the electromagnetic field that are comparable to, or smaller than the atomic spacing.\\
%
%
	 \subsection{Hamiltonian along a classical trajectory}
	 \par
	 As in the vacuum theory of quantum electrodynamics, it will be advantageous to expand our field operators, \((\hat{\boldsymbol{A}}(\boldsymbol{x},t),\hat{\boldsymbol{\Pi}}_{A}(\boldsymbol{x},t),\hat{\boldsymbol{X}}_{\omega}(\boldsymbol{x},t),\hat{\boldsymbol{\Pi}}_{X_{\omega}}(\boldsymbol{x},t),\dots)\) in terms of a certain basis.  This expansion will be carried out in terms of the classical solutions given by  (\ref{simple-osc-sol}) and (\ref{Esol}).  Explicitly, the classical amplitudes \(\boldsymbol{h}_{X_{\omega}}\) \& \(\boldsymbol{h}_{Y_{\omega}}\) that act as sources for the electromagnetic field (e.g. see (\ref{Esol})) will become operators, analogous to the \(\hat{\boldsymbol{a}}(\boldsymbol{k})\) and \(\hat{\boldsymbol{a}}^{\dagger}(\boldsymbol{k})\) of the vacuum theory~\cite{berestetskii_quantum_2004}.  As we will be expanding the space--time dependence of the field operators in terms of these classical solutions, we can use this to simplify the form of the Hamiltonian.  This simplification is equivalent to the usual procedure of Fano diagonalization used in~\cite{huttner__1992,philbin__2010}.
	 \par
	 It turns out that the most straightforward route to simplify the Hamiltonian is to begin with the part associated with the continuum of oscillators, \(H_{\text{\tiny{R}}}\).  After applying (\ref{Xcanmom}), this is
	 \begin{widetext}
	 \begin{equation}
	 	H_{\text{\tiny{R}}}=\int d^{2}\boldsymbol{x}_{\parallel}\int\frac{d k}{2\pi}\int\frac{d\Omega}{2\pi}\int\frac{d\Omega^{\prime}}{2\pi}e^{-i(\Omega+\Omega^{\prime})t}
		\int_{0}^{\infty}d\omega
		(\omega^{2}-\Omega_{+}\Omega_{+}^{\prime})\bigg[\tilde{\boldsymbol{X}}_{\omega}(k,\Omega)\cdot\tilde{\boldsymbol{X}}_{\omega}(-k,\Omega^{\prime})+\tilde{\boldsymbol{Y}}_{\omega}(k,\Omega)\cdot\tilde{\boldsymbol{Y}}_{\omega}(-k,\Omega^{\prime})\bigg]\label{fourier-H-R}
	 \end{equation}
	 \end{widetext}
	 Into (\ref{fourier-H-R}) we insert (\ref{simple-osc-sol}) and its magnetic counterpart, to which we apply the results, \(2\Omega_{-}^{2}-\Omega_{+}\Omega_{+}^{\prime}-\Omega_{-}\Omega_{-}^{\prime}=2\gamma\Omega(\Omega_{-}-\Omega_{+}^{'})\), (\ref{identity-1}), and (\ref{field-G}).  After some lengthy manipulations this leads to
	 \begin{multline}
		H_{\text{\tiny{R}}}=\int d^{2}\boldsymbol{x}_{\parallel}\int \frac{d k}{2\pi}\int_{0}^{\infty}2\omega\omega_{+}\left|\boldsymbol{h}_{X_{\gamma\omega}}(\boldsymbol{x}_{\parallel},k)\right|^{2}d\omega\\
		+(X\to Y)-H_{\text{\tiny{F}}}\label{resham}
	\end{multline}
	where we have introduced \(\omega_{\pm}=\gamma(\omega\pm Vk)\).  The total Hamiltonian is therefore equal to an integration over the absolute squares of the amplitudes, \(\boldsymbol{h}_{X_{\gamma\omega}}\) \& \(\boldsymbol{h}_{Y_{\gamma\omega}}\) multiplied by \(\omega\omega_{+}\), which can be either positive or negative.  As in the previous section, it is clear that for certain choices of the amplitudes, it is possible for this classical Hamiltonian to take an arbitrarily negative value.\\
%
%
	 \section{Quantized Hamiltonian}
	 \par
	 To quantize the Hamiltonian given by the sum, (\ref{resham}) + \(H_{\text{\tiny{F}}}\), we take the usual approach of quantum field theory and let the expansion coefficients become operators.  In this case the expansion coefficients are given by the \(\boldsymbol{h}_{X_{\omega}}\) and \(\boldsymbol{h}_{Y_{\omega}}\), and we perform the following substitution
	 \begin{align}
	 	\boldsymbol{h}_{X_{\gamma\omega}}(\boldsymbol{x}_{\parallel},k)&\to\sqrt{\frac{\hbar}{2\omega}}\hat{\boldsymbol{C}}_{\text{\tiny{E}}}(\boldsymbol{x}_{\parallel},k,\gamma\omega)\nonumber\\
		\boldsymbol{h}_{Y_{\gamma\omega}}(\boldsymbol{x}_{\parallel},k)&\to\sqrt{\frac{\hbar}{2\omega}}\hat{\boldsymbol{C}}_{\text{\tiny{B}}}(\boldsymbol{x}_{\parallel},k,\gamma\omega)\label{quant-class}
	 \end{align}
	 where it is assumed that the \(\hat{\boldsymbol{C}}\) operators are bosonic
	 \begin{multline}
	 	\left[\hat{\boldsymbol{C}}_{\lambda}(k,\boldsymbol{x}_{\parallel},\Omega),\hat{\boldsymbol{C}}^{\dagger}_{\lambda^{\prime}}(k^{\prime},\boldsymbol{x}_{\parallel}^{\prime},\Omega^{\prime})\right]=\\
		2\pi\boldsymbol{\mathbb{1}}_{3}\delta_{\lambda\lambda^{\prime}}\delta(\Omega-\Omega^{\prime})\delta(k-k^{\prime})\delta^{(2)}(\boldsymbol{x}_{\parallel}-\boldsymbol{x}_{\parallel}^{\prime})\label{bosonic}
	 \end{multline}
	 The (normal ordered) Hamiltonian is then
	 \begin{equation}
		\hat{H}=\sum_{\lambda=\text{\tiny{E,B}}}\int d^{2}\boldsymbol{x}_{\parallel}\int \frac{d k}{2\pi}\int_{0}^{\infty}d\omega\hbar\omega_{+}\hat{\boldsymbol{C}}^{\dagger}_{\lambda}(\gamma\omega)\cdot\hat{\boldsymbol{C}}_{\lambda}(\gamma\omega)\label{quantum-hamiltonian}
	\end{equation}
	where for brevity we have suppressed the arguments, \(k\) and \(\boldsymbol{x}_{\parallel}\) within the \(\hat{\boldsymbol{C}}_{\lambda}\) operators.  As expected, the lack of a lower bound for the classical Hamiltonian carries over to the quantum case, the eigenstates of (\ref{quantum-hamiltonian}) having eigenvalues from \(-\infty\) to \(\infty\).
	\par
	It is shown in appendix \ref{app-B} that this specification of operators and commutation relations is consistent with the canonical commutation relations that must hold between the fields and their conjugate momenta.  It might be objected that we have substituted (\ref{quant-class}) in the final classical result (\ref{resham}), which itself assumes that \(\boldsymbol{h}_{X_{\omega}}\) \& \(\boldsymbol{h}_{X_{\omega}}^{\star}\) commute.  However, the Hamiltonian is quadratic in all variables, so in doing this we have only neglected a constant term which is equivalent to an unobservable phase factor in the overall wave function, and can therefore be dropped.\\
%
%
	\section{Detector coupled to the electromagnetic field\label{detector-section}}
	\par
	The Hamiltonian (\ref{quantum-hamiltonian}) is interesting, but at present it is simply that of~\cite{philbin__2010}, described within a different frame of reference.  To demonstrate the physical effects of a relatively moving dielectric, we must couple another system to the electromagnetic field.  As a model we consider a point--like detector at a fixed position, \(\boldsymbol{x}_{0}\), with an internal variable, \(\boldsymbol{x}\), and some internal energy states which are the eigenfunctions of a Hamiltonian, \(\hat{H}_{D}\).
	\par
	The detector (atom) is coupled to the electromagnetic field in the usual form dictated by minimal coupling~\cite{craig_molecular_1998}.  The interaction terms are then transformed into a gauge invariant form through a unitary transformation of the operators, \(\hat{O}=\hat{U}\hat{O}^{\prime}\hat{U}^{\dagger}\) where in the dipole approximation, \(\hat{U}\simeq\exp(\frac{ie}{\hbar}\boldsymbol{x}\boldsymbol{\cdot}\hat{\boldsymbol{A}}(\boldsymbol{x}_{0}))\)~\cite{loudon1983,ackerhalt1984}.  This then gives the following familiar expression for the full (interaction picture) Hamiltonian~\footnote{Due to the unitary transformation, the `electric field operator' in (\ref{detector-ham}) differs from the true electric field operator by a term proportional to the dipole moment~\cite{ackerhalt1984}.  However this difference does not affect our calculation of the transition rate.}
	\begin{equation}
		\hat{H}=\hat{H}_{D}-\hat{\boldsymbol{d}}\boldsymbol{\cdot}\hat{\boldsymbol{E}}(\boldsymbol{x}_{0},t)+\hat{H}_{M+F}\label{detector-ham}
	\end{equation}
	where \(\hat{\boldsymbol{d}}=e\boldsymbol{x}\) is the dipole operator that acts on the internal state of the atom, \(\hat{H}_{M+F}\) is given by (\ref{quantum-hamiltonian}), and the electric field operator, \(\hat{\boldsymbol{E}}(\boldsymbol{x}_{0},t)\), is expanded as in (\ref{op-expansion}), with the coefficients (\ref{E-expansion}).  
	\par
	We consider the Hamiltonian associated with the internal states of the atom to be in the form of equally spaced energy levels, \(\hat{H}_{D}=\hbar\omega\hat{a}^{\dagger}\hat{a}\), and the interaction to be of the form, \(\hat{H}_{I}(t)=-i\boldsymbol{\kappa}\boldsymbol{\cdot}\hat{\boldsymbol{E}}(\boldsymbol{x}_{0},t)(\hat{a}e^{-i\omega t}-\hat{a}^{\dagger}e^{i\omega t})\), where the \(\hat{a}\) \& \(\hat{a}^{\dagger}\) are the usual raising and lowering operators, and \(\boldsymbol{\kappa}\) is a constant vector.  Note that in the presence of a lossy dielectric, the atom is coupled to the full electric field, rather than just the transverse part~\cite{barnett1992,barnett1996}.
	\par
	It is not possible to introduce a set of orthonormal number states to use with the \(\hat{\boldsymbol{C}}\) operators.  In vacuum QED one would introduce periodic boundary conditions, and the dispersion relation would then restrict both the wave--vector and the frequency to discrete values so that a set of such states could be introduced.  In our case the dissipation within the medium removes the dispersion relation between the frequency and the wave--vector for excitations within the medium, and this option is not open to us.  A straightforward alternative is to use a set of orthogonal states with infinite norm, defined over a continuum, as in~\cite{dung2000}.  However, here we use the approach of~\cite{blow__1990}, and work in terms of a new set of operators for which number states can be introduced, labelled by four integers \((l,m,n,p)\),
	\begin{multline*}
		\hat{c}_{\lambda,\sigma}(l,m,n,p)=\int \frac{d k}{2\pi}\int d^{2}\boldsymbol{x}_{\parallel}\int_{0}^{\infty}d\Omega\\
		\times\phi_{l,m,n,p}(k,\boldsymbol{x}_{\parallel}\Omega)\hat{C}_{\lambda,\sigma}(k,\boldsymbol{x}_{\parallel},\Omega)
	\end{multline*}
	where we have expanded the normal mode operators in terms of unit vectors, \(\boldsymbol{e}_{\sigma}\): \(\hat{\boldsymbol{C}}_{\lambda}=\sum_{\sigma}\boldsymbol{e}_{\sigma}\hat{C}_{\lambda,\sigma}\), and chosen the orthonormal basis functions, \(\phi_{l,m,n,p}\)
	\begin{multline*}
		\int \frac{d k}{2\pi}\int d^{2}\boldsymbol{x}_{\parallel}\int_{0}^{\infty}d\Omega\,\phi_{l_{1},m_{1},n_{1},p_{1}}\phi^{\star}_{l_{2},m_{2},n_{2},p_{2}}\\
		=\delta_{l_{1}l_{2}}\delta_{m_{1}m_{2}}\delta_{n_{1}n_{2}}\delta_{p_{1}p_{2}}
	\end{multline*}
	that form a complete set,
	\begin{multline}
		\sum_{l,m,n,p}\phi_{l,m,n,p}(k,\boldsymbol{x}_{\parallel},\Omega)\phi^{\star}_{l,m,n,p}(k^{\prime},\boldsymbol{x}_{\parallel}^{\prime},\Omega^{\prime})\\
		=2\pi\delta(k-k^{\prime})\delta^{(2)}(\boldsymbol{x}_{\parallel}-\boldsymbol{x}^{\prime}_{\parallel})\delta(\Omega-\Omega^{\prime})\label{completeness-relation}
	\end{multline}
	These new operators then satisfy the familiar commutation relations
	\begin{multline*}
		\left[\hat{c}_{\lambda_{1},\sigma_{1}}(l_{1},m_{1},n_{1},p_{1}),\hat{c}_{\lambda_{2},\sigma_{2}}(l_{2},m_{2},n_{2},p_{2})\right]\\
		=\delta_{\lambda_{1}\lambda_{2}}\delta_{\sigma_{1}\sigma_{2}}\delta_{l_{1}l_{2}}\delta_{m_{1}m_{2}}\delta_{n_{1}n_{2}}\delta_{p_{1}p_{2}}
	\end{multline*}
	and we can use this fact to define number states,
	\begin{equation}
		|n_{\lambda,\sigma,i,j,k,l}\rangle=\frac{1}{\sqrt{n!}}\left[\hat{c}_{\lambda,\sigma}^{\dagger}(i,j,k,l)\right]^{n}|0\rangle\label{number-states}
	\end{equation}\\
	If we suppose that the detector is initially in the ground state and the combined system of field and dielectric is in the zero particle state, \(|0\rangle\otimes|0\rangle\), then to first order in \(|\boldsymbol{\kappa}|\) the only possible transition for the detector is into the state \(|1\rangle\), and the field--matter system into a state, \(|1_{\lambda,\sigma,i,j,k,l}\rangle\).  We therefore write the complete wave--function as,
	\[
		|\psi\rangle=|0\rangle\otimes|0\rangle+\sum_{\lambda,\sigma,i,j,k,l}\zeta_{\lambda,\sigma,i,j,k,l}(t)|1\rangle\otimes|1_{\lambda,\sigma,i,j,k,l}\rangle
	\]\\
	The usual results of time--dependent perturbation theory can now be applied~\cite{landau_quantum_2003}.  The rate of change of the quantity \(\zeta_{\lambda,\sigma,i,j,k,l}(t)\) is,	
	\begin{equation}
		\dot{\zeta}_{\lambda,\sigma,i,j,k,l}(t)=-\frac{i}{\hbar}\langle1_{\lambda,\sigma,i,j,k,l}|\otimes\langle1|\hat{H}_{I}(t)|0\rangle\otimes|0\rangle\label{expansion-change}
	\end{equation}
	Inserting the expression for the interaction Hamiltonian, along with the expansion of the electric field in terms of (\ref{E-expansion}) and integrating over a time interval, \([0,T]\), gives the amplitude for excitation after a time interval, \(T\), \(\zeta_{\lambda,\sigma,i,j,k,l}(T)\).  The rate of excitation of the detector, \(\Gamma_{0\to1}\) is given by the absolute square of this amplitude summed over all possible indices, and divided by the time interval, \(T\).  We take \(T\) to infinity, and obtain the result that would be expected from an application of Fermi's golden rule
	\begin{equation}
		\Gamma_{0\to1}=\frac{2\omega^{2}}{\hbar}\int_{\omega/V}^{\infty}\boldsymbol{\kappa}\boldsymbol{\cdot}\text{Im}\left[\boldsymbol{G}(-k,\boldsymbol{x}_{\parallel 0},\boldsymbol{x}_{\parallel 0},-\omega)\right]\boldsymbol{\cdot}\boldsymbol{\kappa}\,\frac{d k}{2\pi}\label{transition-rate}
	\end{equation}
	Full details of the derivation of this result can be found in appendix \ref{appendix-D}.  In general the transition rate (\ref{transition-rate}) is non--zero, and it is therefore possible to extract energy from the centre of mass motion of a moving dielectric through coupling to the electromagnetic field within the dielectric, or in its vicinity~\footnote{In the case of a detector embedded in a moving dielectric one must introduce a correction factor to account for the fact that the dipole moment couples to the local microscopic electric field, rather than the macroscopic one.  To obtain a finite emission or absorption rate, one must also find a way to regularize the coupling to the longitudinal part of the electric field.  A thorough discussion of these issues is given within~\cite{barnett1996}.}.  This is a physical effect of the negative energy states within (\ref{quantum-hamiltonian}) that arises from our description of dissipation.
	\par
	Following the concluding discussion of section~\ref{classical-hamiltonian}, it is clear that for moderate velocities we can only rely on (\ref{transition-rate}) for low frequency excitations of the detector.  For instance, if we were to consider a dielectric moving at \(1 \text{ms}^{-1}\) then an optical excitation \(\omega\sim10^{15} \text{Hz}\) would correspond to integrating over wave--vectors, \(k>10^{15}\text{m}^{-1}\) which is clearly far beyond a regime where the behaviour of the dielectric can be described with a susceptibility like (\ref{susdef}).  If the integrand in (\ref{transition-rate}) decays rapidly (as it does in (\ref{transition-rate-surface})), then an \(\omega\) in the GHz regime (\(k\sim10^{9}\)) appears to be more appropriate.  For relativistic velocities, (\ref{transition-rate}) can be applied to much higher frequency excitations.  Usually such velocities are completely inaccessible to experiment.  However, they are worth considering in this case, for in experiments such as~\cite{belgiorno__2010}, a refractive index perturbation is created that moves through a dielectric at a fraction of the speed of light in vacuum.  In the rest frame of the perturbation we have the dielectric moving at a uniform relativistic velocity that is subject to a weak stationary perturbation.  It is therefore likely that the experimentally observed production of photons can be understood with some analogous formula to (\ref{transition-rate}).  This will be the subject of future work.
	\par
	The transition rate (\ref{transition-rate}) vanishes when \(V=0\).  This is because the argument of the delta function (\ref{delta-identity}) in the derivation of (\ref{transition-rate}) is never zero when \(V=0\) and so \(\Gamma_{0\to1}=0\).  However---consistent with the concluding discussion of section \ref{classical-hamiltonian}---the way it tends to zero depends on the response of the medium to large wave-vectors at a fixed frequency.  If the atom is embedded within a dielectric medium this requires knowledge of the spatial dispersion of the susceptibilities.  Yet when the atom is in a region of free space outside of a moving dielectric medium (see section \ref{surface-section} below), this is not necessary as the large wave-vector modes are very tightly bound to the surface.  Therefore in this case the transition rate (\ref{transition-rate-surface}) falls to zero as \(V\to0\) because the relevant modes do not reach the atom (see (\ref{transition-rate-surface})).
	\par
	We now investigate (\ref{transition-rate}) in two simple cases; free space, and close to a surface.
%
%
	\subsubsection{Free space}
	\begin{figure}
		\includegraphics[width=8cm]{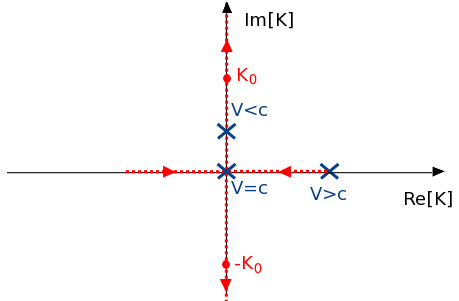}
		\caption{As \(k\) is increased within the integrand of (\ref{transition-rate}), the poles at \(\pm K_{0}\) in (\ref{free-green-integral}) move along the red dotted lines in the directions indicated by red arrows.  When \(V<c\), \(K_{0}\) starts at a point on the imaginary axis indicated by a blue cross and moves vertically upwards.  In the case when \(V=c\), \(K_{0}\) starts at the origin.  Therefore, for sub--luminal propagation, the poles never touch the real axis.  Meanwhile, for super--luminal propagation, \(K_{0}\) starts at some point on the real axis, moves towards the origin, and then up the imaginary axis.  In this case there is an imaginary contribution to the integral along positive \(\text{Re}[K]\) in (\ref{free-green-integral}) that comes from the need to negotiate the pole at \(K_{0}\).\label{figure-2}}
	\end{figure}
	Firstly, we show that \(\Gamma_{0\to1}\) vanishes in free space.  The free space Green function is a function of the difference in the positions, \(\boldsymbol{x}\) \& \(\boldsymbol{x}^{\prime}\), and in \(\boldsymbol{k}\) space is given by,
	\begin{equation}
		\boldsymbol{G}(\boldsymbol{k},\omega)=-\mu_{0}\frac{\boldsymbol{k}\boldsymbol{\otimes}\boldsymbol{k}-(\omega/c)^{2}\boldsymbol{\mathbb{1}}_{3}}{(\omega/c)^{2}(|\boldsymbol{k}|-\omega/c-i\eta)(|\boldsymbol{k}|+\omega/c+i\eta)}\label{free-space-G}
	\end{equation}
	We transform this into the \(y\)--\(z\) plane of physical space, \(\boldsymbol{G}(k,\boldsymbol{x}_{\parallel}-\boldsymbol{x}_{\parallel}^{\prime},\omega)=(2\pi)^{-2}\int \boldsymbol{G}(k,\boldsymbol{K},\omega) \exp[i\boldsymbol{K}\boldsymbol{\cdot}(\boldsymbol{x}_{\parallel}-\boldsymbol{x}_{\parallel}^{\prime})]d^{2}\boldsymbol{K}\), integrate over the polar angle in the \(\boldsymbol{K}\) plane, and find the result,
	\begin{multline}
		\text{Im}\left[\boldsymbol{G}(k,\boldsymbol{0},\omega)\right]=\mu_{0}\left[\frac{c^{2}}{\omega^{2}}(-k^{2}\hat{\boldsymbol{x}}\boldsymbol{\otimes}\hat{\boldsymbol{x}}+\boldsymbol{\nabla}_{\parallel}\boldsymbol{\otimes}\boldsymbol{\nabla}_{\parallel})+\boldsymbol{\mathbb{1}}_{3}\right]\\
		\times\text{Im}\left[\int_{0}^{\infty}\frac{K dK}{2\pi}\frac{J_{0}(K|\boldsymbol{x}_{\parallel}-\boldsymbol{x}_{\parallel}^{\prime}|)}{(K-K_{0}-i\eta)(K+K_{0}+i\eta)}\right]_{\boldsymbol{x}_{\parallel}\to\boldsymbol{x}_{\parallel}^{\prime}}\label{free-green-integral}
	\end{multline}
	where \(K_{0}=\sqrt{(\omega/c)^{2}-k^{2}}\), \(J_{0}\) is a zeroth order Bessel function, and \(\boldsymbol{\nabla}_{\parallel}=\partial/\partial\boldsymbol{x}_{\parallel}\).  Terms involving first order derivatives in \(\boldsymbol{x}_{\parallel}\) have been set to zero given that they serve to replace the zeroth order Bessel function with a first order Bessel function, which vanishes as \(\boldsymbol{x}_{\parallel}\to\boldsymbol{x}_{\parallel}^{\prime}\).  The only possible contribution to the imaginary part of the integral in (\ref{free-green-integral}) would come from the need to negotiate the poles if they meet the real line as \(\eta\to0\).  However, when \(k>\omega/c\) then \(K_{0}\) is imaginary (see figure \ref{figure-2}).  In this case these poles never meet the real line, and the right hand side of (\ref{free-green-integral}) vanishes.  Therefore, so long as \(V<c\) then (\ref{transition-rate}) is zero in free space.  This is consistent with existing findings that the vacuum is unstable to superluminal propagation (e.g. ~\cite{cohen2011}).
%
%
	\subsubsection{Close to a surface\label{surface-section}}
	\begin{figure}
		\includegraphics[width=5cm]{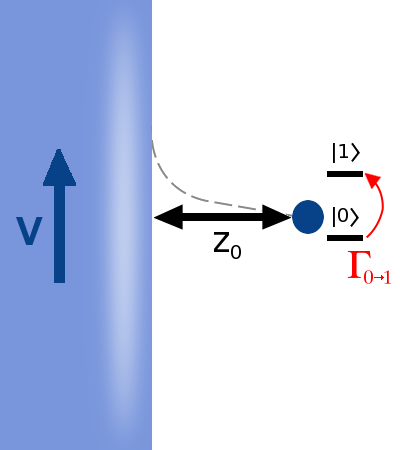}
		\caption{We consider the transition rate, \(\Gamma_{0\to1}\) of a detector from the ground state, \(|0\rangle\) to an excited state \(|1\rangle\) as a function of the perpendicular distance, \(z_{0}\) from a surface moving along the \(x\) axis at \(T=0 \text{K}\).  The rate of excitation is given by (\ref{transition-rate-surface}), and decays with \(z_{0}\) at some rate (grey dashed line) determined by an integration over the imaginary parts of the reflection coefficients of the moving surface.\label{fig-3}}
	\end{figure}
	\par
	We now consider the possibility of exciting an internal state of a system through interaction with the vacuum electromagnetic field in the region of space outside of a moving dielectric.  The equivalent case of atoms moving along surfaces has been investigated before (e.g.~\cite{scheel2009,barton2010b}), but usually the work is concerned with the drag force on the atom due to the vacuum field.  The author is not aware of equation (\ref{transition-rate}) being derived previously, or being applied to the internal states of quantum systems moving over surfaces.
	\par
	For a detector situated outside of a moving dielectric (see figure \ref{fig-3}), we can obtain the expression for the Green function through taking (\ref{free-space-G}) and writing it in the following form,
	\begin{multline}
		\boldsymbol{G}(k,\boldsymbol{x}_{\parallel},\boldsymbol{x}_{\parallel}^{\prime},\omega)=\frac{\mu_{0}c^{2}}{\omega^{2}}\int\frac{d^{2}\boldsymbol{K}}{(2\pi)^{2}}e^{i\boldsymbol{K}\boldsymbol{\cdot}(\boldsymbol{x}_{\parallel}-\boldsymbol{x}_{\parallel}^{\prime})}\\
		\times\bigg[\frac{|\boldsymbol{k}|^{2}\sum_{\lambda}\hat{\boldsymbol{e}}_{\lambda}(\boldsymbol{k})\boldsymbol{\otimes}\hat{\boldsymbol{e}}_{\lambda}(\boldsymbol{k})}{(|\boldsymbol{k}|-\omega/c-i\eta)(|\boldsymbol{k}|+\omega/c+i\eta)}-\boldsymbol{\mathbb{1}}_{3}\bigg]\label{green-function-moving-surface}
	\end{multline}
	where the \(\hat{\boldsymbol{e}}_{1,2}\) represent the two unit vectors orthogonal to \(\hat{\boldsymbol{k}}\), the unit wave-vector.  We take a dielectric moving along \(\hat{\boldsymbol{x}}\) and lying in the \(x\)--\(y\) plane with free space in the region \(z>0\).   The detector is positioned near to the surface (\(z_0>0\)).  In this case when \(\boldsymbol{x}^{\prime}\) is set as the detector position, only the first term within the integrand of (\ref{green-function-moving-surface}) is non zero at the surface.  In satisfying the boundary conditions at the interface we therefore ignore the term proportional to \(\boldsymbol{\mathbb{1}}_{3}\).  Performing one of the \(\boldsymbol{K}\) integrals (say over \(k_{z}\)) enforces the dispersion relation in the remaining term, and (\ref{green-function-moving-surface}) becomes an integral over incident propagating waves.  The boundary conditions at \(z=0\) can then be fulfilled with the usual reflection coefficients, \(r_{\lambda\lambda^{\prime}}(\boldsymbol{k},\omega)\), where \(\lambda\) and \(\lambda^{\prime}\) each take the values \(1\) or \(2\), with the meaning that polarization \(\lambda\) is incident, and polarization \(\lambda^{\prime}\) is reflected (moving media mix polarizations~\cite{horsley2011b})
	\begin{widetext}
	\begin{multline}
	\boldsymbol{G}(k,\boldsymbol{x}_{\parallel},\boldsymbol{x}_{\parallel0},\omega)=\mu_{0}\bigg\{\int_{-\infty}^{\infty}\frac{d k_y}{2\pi}\frac{e^{i k_{y}(y-y_{0})}}{2\xi}\bigg[\sum_{\lambda}\hat{\boldsymbol{e}}_{\lambda}(k,k_{y},is\xi)\boldsymbol{\otimes}\hat{\boldsymbol{e}}_{\lambda}(k,k_{y},is\xi)e^{-s\xi(z-z_{0})}\\
	+\sum_{\lambda\lambda^{\prime}}\hat{\boldsymbol{e}}_{\lambda^{\prime}}(k,k_{y},i\xi)\boldsymbol{\otimes}\hat{\boldsymbol{e}}_{\lambda}(k,k_{y},-i\xi)r_{\lambda\lambda^{\prime}}(k,k_{y},-i\xi,\omega)e^{-\xi(z+z_{0})}\bigg]-\frac{c^{2}}{\omega^{2}}\boldsymbol{\mathbb{1}}_{3}\delta^{(2)}(\boldsymbol{x}_{\parallel}-\boldsymbol{x}_{\parallel0})\bigg\}\label{moving-surface-G}
	\end{multline}
	where \(\xi=\sqrt{k^{2}+k_{y}^{2}-\omega^{2}/c^{2}}\) and \(s=\text{sign}(z-z_{0})\).  At \(\boldsymbol{x}_{\parallel}=\boldsymbol{x}_{\parallel0}\), the first term in the integrand of (\ref{moving-surface-G}) becomes real due to quantities linear in \(s\) going to zero.  Applying (\ref{transition-rate}) to (\ref{moving-surface-G}) then gives,
	\begin{align}
		\Gamma_{0\to1}&=\frac{\mu_{0}\omega^{2}}{\hbar}\sum_{\lambda,\lambda^{\prime}}\int_{\omega/V}^{\infty}\frac{d k}{2\pi}\int_{-\infty}^{\infty}\frac{d k_{y}}{2\pi\xi}e^{-2\xi z_{0}}\text{Im}\bigg[\boldsymbol{\kappa}\boldsymbol{\cdot}\hat{\boldsymbol{e}}_{\lambda^{\prime}}(-k,k_{y},i\xi)\boldsymbol{\kappa}\boldsymbol{\cdot}\hat{\boldsymbol{e}}_{\lambda}(-k,k_{y},-i\xi)r_{\lambda\lambda^{\prime}}(-k,k_{y},-i\xi,-\omega)\bigg]\label{transition-rate-surface}\\
		&\sim\frac{\mu_{0}\omega^{2}}{\hbar}\sum_{\lambda}\int_{\omega/V}^{\infty}\frac{d k}{2\pi}\int_{-\infty}^{\infty}\frac{d k_{y}}{2\pi\xi}e^{-2\xi z_{0}}\left|\boldsymbol{\kappa}\boldsymbol{\cdot}\hat{\boldsymbol{e}}_{\lambda}(-k,k_{y},i\xi)\right|^{2}\text{Im}\bigg[r_{\lambda\lambda}(-k,k_{y},-i\xi,-\omega)\bigg]\label{approx-trans}
	\end{align}
	\end{widetext}
	where the second line is obtained through neglecting the mixed reflection coefficients, \(r_{12}\) \& \(r_{21}\), which are a factor of \(V/c\) smaller than \(r_{11}\) and \(r_{22}\).  Something like expression (\ref{approx-trans}) could have been anticipated from applying the reasoning of~\cite{pendry__1997} which contains very similar expressions in the discussion of the phenomenon of quantum friction, based not on canonical quantization but on the fluctuation--dissipation theorem.  For positive rest frame frequencies, \(\omega_{-}>0\) the imaginary part of the reflection coefficient is positive.  The range of the integration over \(k\) ensures that this is always the case.  As the detector is moved away from the surface, the exponential decay on the right of (\ref{transition-rate-surface}) serves to reduce \(\Gamma_{0\to1}\) to zero.  However, the dependence of this rate on the distance from the dielectric depends on the dispersion of the medium.  This will be investigated in future work.
%
%
	\section{Conclusions}
	\par
	We have developed a formalism for the description of uniformly moving media that includes a full account of dispersion and dissipation.  It was found that due to the presence of dissipation, the Hamiltonian describing the interaction between a uniformly moving dielectric and the electromagnetic field lacks a lower bound (although this statement should be qualified as in section \ref{classical-hamiltonian}).  Further to this it was shown that a stationary detector placed inside or outside a moving medium has a finite probability of excitation that depends on the specific dispersion of the medium, and in the case of being outside the medium, the distance from the surface. 
	\acknowledgements
	The author thanks Tom Philbin, Maurizio Artoni, Giuseppe La Rocca, Steve Barnett, Daniele Faccio, and Daniel Oi for useful discussions.  Thanks to the referee, who introduced me to~\cite{dung2000} and whose comments greatly improved the manuscript.  This work was supported by the EPSRC.
%
%
	 \begin{widetext}
	 \appendix
%
%
	 \section{Integral identity for susceptibilities}
	 \par
	 To obtain expression (\ref{resham}) the following result must be applied,
	  \begin{multline}
	 	\int_{0}^{\infty}d\omega\frac{(\omega^{2}-\Omega_{+}\Omega_{+}^{\prime})[\boldsymbol{\alpha_{\text{\tiny{EE}}}}(\omega)\boldsymbol{\cdot}\boldsymbol{\alpha_{\text{\tiny{EE}}}^{\text{\tiny{T}}}}(\omega)+\boldsymbol{\alpha_{\text{\tiny{EB}}}}(\omega)\boldsymbol{\cdot}\boldsymbol{\alpha^{\text{\tiny{T}}}_{\text{\tiny{EB}}}}(\omega)]}{(\omega-\Omega_{-}-i\eta)(\omega+\Omega_{-}+i\eta)(\omega-\Omega_{+}^{\prime}-i\eta)(\omega+\Omega_{+}^{\prime}+i\eta)}=\frac{(\Omega_{-}^{2}-\Omega_{+}\Omega_{+}^{\prime})\boldsymbol{\chi}_{\text{\tiny{EE}}}(\Omega_{-})}{(\Omega_{-}-\Omega_{+}^{\prime}-i\eta)(\Omega_{-}+\Omega_{+}^{\prime}+i\eta)}\\
		+\frac{({\Omega^{\prime}_{+}}^{2}-\Omega_{+}\Omega_{+}^{\prime})\boldsymbol{\chi}_{\text{\tiny{EE}}}(\Omega_{+}^{\prime})}{(\Omega_{+}^{\prime}-\Omega_{-}-i\eta)(\Omega_{+}^{\prime}+\Omega_{-}+i\eta)}\label{identity-1}
	 \end{multline}
	 along with similar expressions containing the susceptibilities, \(\boldsymbol{\chi}_{\text{\tiny{EB}}}\),  \(\boldsymbol{\chi}_{\text{\tiny{BE}}}\), and \(\boldsymbol{\chi}_{\text{\tiny{BB}}}\).  The proof of (\ref{identity-1}) is similar to that of the usual Kramers--Kronig relations.  We first expand the left hand side of (\ref{identity-1}) in terms of residues at poles and a principal part,
	  \begin{multline}
	 	\int_{0}^{\infty}d\omega\frac{(\omega^{2}-\Omega_{+}\Omega_{+}^{\prime})[\boldsymbol{\alpha_{\text{\tiny{EE}}}}(\omega)\boldsymbol{\cdot}\boldsymbol{\alpha_{\text{\tiny{EE}}}^{\text{\tiny{T}}}}(\omega)+\boldsymbol{\alpha_{\text{\tiny{EB}}}}(\omega)\boldsymbol{\cdot}\boldsymbol{\alpha^{\text{\tiny{T}}}_{\text{\tiny{EB}}}}(\omega)]}{(\omega-\Omega_{-}-i\eta)(\omega+\Omega_{-}+i\eta)(\omega-\Omega_{+}^{\prime}-i\eta)(\omega+\Omega_{+}^{\prime}+i\eta)}=\\
		i\frac{(\Omega_{-}^{2}-\Omega_{+}\Omega_{+}^{\prime})\text{Im}[\boldsymbol{\chi}_{\text{\tiny{EE}}}(\Omega_{-})]}{(\Omega_{-}-\Omega_{+}^{\prime}-i\eta)(\Omega_{-}+\Omega_{+}^{\prime}+i\eta)}+i\frac{({\Omega_{+}^{\prime}}^{2}-\Omega_{+}\Omega_{+}^{\prime})\text{Im}[\boldsymbol{\chi}_{\text{\tiny{EE}}}(\Omega_{+}^{\prime})]}{(\Omega_{+}^{\prime}-\Omega_{-}-i\eta)(\Omega_{+}^{\prime}+\Omega_{-}+i\eta)}\\
		+\frac{2}{\pi}\text{P}\int_{0}^{\infty}d\omega\frac{\omega(\omega^{2}-\Omega_{+}\Omega_{+}^{\prime})\text{Im}[\boldsymbol{\chi}_{\text{\tiny{EE}}}(\omega)]}{(\omega^{2}-\Omega_{-}^{2})(\omega^{2}-{\Omega_{+}^{\prime}}^{2})}\label{res1}
	 \end{multline}
	 where we have applied the result (see (\ref{susdef})) \(\boldsymbol{\alpha_{\text{\tiny{EE}}}}\boldsymbol{\cdot}\boldsymbol{\alpha_{\text{\tiny{EE}}}^{\text{\tiny{T}}}}+\boldsymbol{\alpha_{\text{\tiny{EB}}}}\boldsymbol{\cdot}\boldsymbol{\alpha^{\text{\tiny{T}}}_{\text{\tiny{EB}}}}=2\omega\text{Im}[\boldsymbol{\chi}_{\text{\tiny{EE}}}]/\pi\).
	 \par
	 Now consider the following integral of a function, \(f(z)\), that is analytic within a closed contour \(\mathcal{C}\),
	 \begin{equation}
	 	\frac{1}{2\pi i}\oint_{\mathcal{C}}\frac{(z^{2}-a)f(z) dz}{(z-z_{1})(z-z_{2})}=\frac{(z_{1}^{2}-a)f(z_{1})-(z_{2}^{2}-a)f(z_{2})}{(z_{1}-z_{2})}\label{conint}
	 \end{equation}
	 where \(a\) is real.  If \(f(z)\) is analytic throughout the upper half complex plane, then \(\mathcal{C}\) may be taken as the real line plus a semicircle at infinity, directed anticlockwise.  If it is additionally supposed that \(f(z)\) is such that the integral over this infinite semicircle is zero (it goes to zero faster than \(1/R\), where \(R\) is the radius of the infinite semicircle), then (\ref{conint}) may be written as,
	 \begin{equation}
	 	 \frac{1}{2\pi i}\int_{-\infty}^{\infty}\frac{(x^{2}-a)f(z) dx}{(x-x_{1}-i\eta)(x-x_{2}-i\eta)}
		 =\frac{(x_{1}^{2}-a)f(x_{1})-(x_{2}^{2}-a)f(x_{2})}{(x_{1}-x_{2})}\label{res2}
	 \end{equation}
	 where \(x_{1}\) \& \(x_{2}\) are real, and \(\eta\) is as in the main text, an infinitesimal quantity that serves to shift the poles away from the real line and into the upper half plane.  Expanding the left hand side of (\ref{res2}) into residues and a principal part
	 \begin{equation}
	 	\frac{1}{\pi i}\text{P}\int_{-\infty}^{\infty}\frac{(x^{2}-a)f(x) dx}{(x-x_{1})(x-x_{2})}
		 =\frac{(x_{1}^{2}-a)f(x_{1})-(x_{2}^{2}-a)f(x_{2})}{(x_{1}-x_{2})}
	 \end{equation}
	 we can obtain the following relation between the real and imaginary parts of \(f(z)\),
	 \begin{equation}
		\frac{2}{\pi}\text{P}\int_{0}^{\infty}\frac{x(x^{2}-a)\text{Im}[f(x)] dx}{(x^{2}-x_{1}^{2})(x^{2}-x_{2}^{2})}
		=\frac{(x_{1}^{2}-a)\text{Re}[f(x_{1})]-(x_{2}^{2}-a)\text{Re}[f(x_{2})]}{(x_{1}^{2}-x_{2}^{2})}\label{res3}	 
	 \end{equation}
	 when (\ref{res3}) is applied to (\ref{res1}) we obtain (\ref{identity-1}).
%
%
	 \section{Field commutation relations\label{app-B}}
	 \par
	 Here we show that the substitution given in (\ref{quant-class}), plus the assumed commutation relations between \(\hat{\boldsymbol{C}}\) \& \(\boldsymbol{C}^{\dagger}\) (\ref{bosonic}) are consistent with the commutation relations that must hold between the canonical momenta and the field amplitudes.  In the gauge where \(\boldsymbol{\nabla}\boldsymbol{\cdot}\boldsymbol{A}=0\), these canonical commutation relations are given by~\cite{loudon1983},
	 \begin{align}
	 \left[\hat{\boldsymbol{A}}(\boldsymbol{x},t),\hat{\boldsymbol{\Pi}}_{\boldsymbol{A}}(\boldsymbol{x}^{\prime},t)\right]&=i\hbar\boldsymbol{\delta}_{\perp}(\boldsymbol{x}-\boldsymbol{x}^{\prime})\nonumber\\
	 \left[\hat{\boldsymbol{X}}_{\omega}(\boldsymbol{x},t),\hat{\boldsymbol{\Pi}}_{X_{\omega^{\prime}}}(\boldsymbol{x}^{\prime},t)\right]&=i\hbar\boldsymbol{\mathbb{1}}_{3}\delta(\omega-\omega^{\prime})\delta^{(3)}(\boldsymbol{x}-\boldsymbol{x}^{\prime})\nonumber\\
	 \left[\hat{\boldsymbol{Y}}_{\omega}(\boldsymbol{x},t),\hat{\boldsymbol{\Pi}}_{Y_{\omega^{\prime}}}(\boldsymbol{x}^{\prime},t)\right]&=i\hbar\boldsymbol{\mathbb{1}}_{3}\delta(\omega-\omega^{\prime})\delta^{(3)}(\boldsymbol{x}-\boldsymbol{x}^{\prime})\label{comm-rel}
	 \end{align}
	 where \(\boldsymbol{\delta}_{\perp}(\boldsymbol{x}-\boldsymbol{x}^{\prime})\) is the transverse delta function,
	 \[
	 	\boldsymbol{\delta}_{\perp}(\boldsymbol{x}-\boldsymbol{x}^{\prime})=\int\frac{d^{3}\boldsymbol{k}}{(2\pi)^{3}\boldsymbol{k}^{2}}\left(\boldsymbol{k}^{2}\boldsymbol{\mathbb{1}}_{3}-\boldsymbol{k}\boldsymbol{\otimes}\boldsymbol{k}\right)e^{i\boldsymbol{k}\boldsymbol{\cdot}(\boldsymbol{x}-\boldsymbol{x}^{\prime})}
	 \]
	 The vector potential operator is expanded in terms of the \(\hat{\boldsymbol{C}}_{\lambda}\) operators as follows,
	 \begin{align}
	 	\hat{\boldsymbol{A}}(\boldsymbol{x},t)&=\sum_{\lambda}\int\frac{d k}{2\pi}\int d^{2}\boldsymbol{x}^{\prime}_{\parallel}\int_{0}^{\infty}d\Omega\sqrt{\frac{\hbar}{2\Omega}}\left[\boldsymbol{f}^{\lambda}_{A}(k,\boldsymbol{x}_{\parallel},\boldsymbol{x}^{\prime}_{\parallel},\Omega)\boldsymbol{\cdot}\hat{\boldsymbol{C}}_{\lambda}(k,\boldsymbol{x}^{\prime}_{\parallel},\gamma\Omega)e^{i(k x-(\Omega+V k)t)}+\text{h.c.}\right]\label{op-expansion}
	 \end{align}
	 with similar expressions holding for the remaining field operators in (\ref{comm-rel}).  To find these expansion coefficients, one must solve the classical problem of section \ref{clas-sol-sec} and then perform the substitution (\ref{quant-class}).  Inserting the operators in the form given by (\ref{op-expansion}) into (\ref{comm-rel}) we obtain the conditions that the expansion coefficients must satisfy in order that the canonical commutation relations be satisfied,
	 \begin{align}
		 \sum_{\lambda}\int\frac{dk}{2\pi}\int d^{2}\boldsymbol{x}_{\parallel1}\int_{-\infty}^{\infty}\frac{d\Omega}{2\gamma\Omega}e^{ik(x-x^{\prime})}\boldsymbol{f}^{\lambda}_{A}(k,\boldsymbol{x}_\parallel,\boldsymbol{x}_{\parallel 1},\Omega)\boldsymbol{\cdot}\boldsymbol{f}^{\lambda\,\dagger}_{\Pi_{A}}(k,\boldsymbol{x}_{\parallel}^{\prime},\boldsymbol{x}_{\parallel1},\Omega)&=i\boldsymbol{\delta}_{\perp}(\boldsymbol{x}-\boldsymbol{x}^{\prime})\label{EM-integral}\\
	 	\sum_{\lambda}\int\frac{dk}{2\pi}\int d^{2}\boldsymbol{x}_{\parallel1}\int_{0}^{\infty}\frac{d\Omega}{2\gamma\Omega}e^{ik(x-x^{\prime})}\bigg[\boldsymbol{f}^{\lambda}_{X_{\omega}}(k,\boldsymbol{x}_\parallel,\boldsymbol{x}_{\parallel 1},\Omega)\boldsymbol{\cdot}\boldsymbol{f}^{\lambda\,\dagger}_{\Pi_{X_{\omega^{\prime}}}}(k,\boldsymbol{x}_{\parallel}^{\prime},\boldsymbol{x}_{\parallel1},\Omega)-\text{c.c.}\bigg]&=i\boldsymbol{\mathbb{1}}_{3}\delta(\omega-\omega^{\prime})\delta^{(3)}(\boldsymbol{x}-\boldsymbol{x}^{\prime})\label{X-integral}
	 \end{align}
	 where we have applied, \(\boldsymbol{f}^{\lambda}_{A}(-k,\boldsymbol{x}_{\parallel},\boldsymbol{x}^{\prime}_{\parallel},-\Omega)=\boldsymbol{f}^{\lambda\star}_{A}(k,\boldsymbol{x}_{\parallel},\boldsymbol{x}^{\prime}_{\parallel},\Omega)\) and a similar relation for \(\boldsymbol{f}^{\lambda}_{\Pi_{A}}(k,\boldsymbol{x}_{\parallel},\boldsymbol{x}^{\prime}_{\parallel},\Omega)\).  In this case the expansion coefficients for the electromagnetic field operators are given by,
	 \begin{align}
	 	\boldsymbol{f}^{\lambda}_{A}(k,\boldsymbol{x}_\parallel,\boldsymbol{x}_{\parallel 1},\Omega)&=\boldsymbol{G}_{\perp}(k,\boldsymbol{x}_{\parallel},\boldsymbol{x}_{\parallel 1},\Omega+Vk)\boldsymbol{\cdot}\hat{\boldsymbol{O}}_{\lambda}^{\dagger}(k,\boldsymbol{x}_{\parallel1},\Omega)\label{A-expansion}\\
		\boldsymbol{f}^{\lambda}_{E}(k,\boldsymbol{x}_\parallel,\boldsymbol{x}_{\parallel 1},\Omega)&=i(\Omega+V k)\boldsymbol{G}(k,\boldsymbol{x}_{\parallel},\boldsymbol{x}_{\parallel 1},\Omega+Vk)\boldsymbol{\cdot}\hat{\boldsymbol{O}}_{\lambda}^{\dagger}(k,\boldsymbol{x}_{\parallel1},\Omega)\label{E-expansion}
	 \end{align}
	 The electric field expansion coefficients (\ref{E-expansion}) are here given for reference. and in the above we define
	 \begin{align}
	 	\hat{\boldsymbol{O}}_{\lambda}(k,\boldsymbol{x}_{\parallel},\Omega)&=[i(\Omega+Vk)\boldsymbol{\alpha}_{\text{\tiny{E$\lambda$}}}^{T}(\boldsymbol{x}_{\parallel},\gamma\Omega)+\boldsymbol{\alpha}_{\text{\tiny{B$\lambda$}}}^{T}(\boldsymbol{x}_{\parallel},\gamma\Omega)\boldsymbol{\cdot}\boldsymbol{\nabla}\boldsymbol{\times}]\nonumber\\
		\hat{\boldsymbol{O}}_{\lambda}^{\dagger}(k,\boldsymbol{x}_{\parallel},\Omega)&=[-i(\Omega+Vk)\boldsymbol{\alpha}_{\text{\tiny{E$\lambda$}}}(\boldsymbol{x}_{\parallel},\gamma\Omega)+\boldsymbol{\times}\stackrel{\leftarrow}{\boldsymbol{\nabla}}^{\star}\boldsymbol{\cdot}\boldsymbol{\alpha}_{\text{\tiny{B$\lambda$}}}(\boldsymbol{x}_{\parallel},\gamma\Omega)]\label{shorthand-operators}
	\end{align} 
	where, as in the main text we have \(\boldsymbol{\nabla}=i k\hat{\boldsymbol{x}}+\boldsymbol{\nabla}_{\parallel}\) and the operation, \(\boldsymbol{G}\boldsymbol{\times}\stackrel{\leftarrow}{\boldsymbol{\nabla}}\) is equivalent to the curl with respect to the right hand index: \(\boldsymbol{G}\boldsymbol{\times}\stackrel{\leftarrow}{\boldsymbol{\nabla}}\equiv e_{ijk}\partial_{j}G_{lk}\).  The Green function within the above formulae is a solution of
	\begin{multline}
		\boldsymbol{\nabla}\boldsymbol{\times}\left[\boldsymbol{\mu}^{-1}(\boldsymbol{x}_{\parallel},\Omega_{-})\boldsymbol{\cdot}\boldsymbol{\nabla}\boldsymbol{\times}\boldsymbol{G}(k,\boldsymbol{x}_{\parallel},\boldsymbol{x}^{\prime}_{\parallel},\Omega)\right]-\Omega^{2}\boldsymbol{\epsilon}(\boldsymbol{x}_{\parallel},\Omega_{-})\boldsymbol{\cdot}\boldsymbol{G}(k,\boldsymbol{x}_{\parallel},\boldsymbol{x}^{\prime}_{\parallel},\Omega)\\
		+i\Omega[\boldsymbol{\chi}_{\text{\tiny{EB}}}(\boldsymbol{x}_{\parallel},\Omega_{-})\boldsymbol{\cdot}\boldsymbol{\nabla}\boldsymbol{\times}\boldsymbol{G}(k,\boldsymbol{x}_{\parallel},\boldsymbol{x}^{\prime}_{\parallel},\Omega)-\boldsymbol{\nabla}\boldsymbol{\times}(\boldsymbol{\chi}_{\text{\tiny{BE}}}(\boldsymbol{x}_{\parallel},\Omega_{-})\boldsymbol{\cdot}\boldsymbol{G}(k,\boldsymbol{x}_{\parallel},\boldsymbol{x}^{\prime}_{\parallel},\Omega))]=\mathbb{1}_{3}\delta^{(2)}(\boldsymbol{x}_{\parallel}-\boldsymbol{x}_{\parallel}^{\prime})\label{field-G}
	\end{multline}
	The susceptibilities are defined as in (\ref{susdef}), with \(\boldsymbol{\epsilon}(\boldsymbol{x}_{\parallel},\Omega_{-})=\epsilon_{0}\boldsymbol{\mathbb{1}}_{3}+\boldsymbol{\chi}_{\text{\tiny{EE}}}(\boldsymbol{x}_{\parallel},\Omega_{-})\), and \(\boldsymbol{\mu}^{-1}(\boldsymbol{x}_{\parallel},\Omega_{-})=\mu_{0}^{-1}\boldsymbol{\mathbb{1}}_{3}-\boldsymbol{\chi}_{\text{\tiny{BB}}}(\boldsymbol{x}_{\parallel},\Omega_{-})\).  We note that in a moving medium the Green function does not satisfy the reciprocity condition,
	 \begin{equation}
	 	\boldsymbol{G}(\boldsymbol{x},\boldsymbol{x}^{\prime},\Omega)\neq\boldsymbol{G}^{T}(\boldsymbol{x}^{\prime},\boldsymbol{x},\Omega)\label{reciprocity}
	\end{equation}
	and therefore \(\boldsymbol{G}^{T}\) does not satisfy (\ref{field-G}) with respect to its second argument, as is usually the case.  Instead it satisfies (\ref{field-G}) with \(\boldsymbol{V}\to-\boldsymbol{V}\) (\(\boldsymbol{\chi}_{\text{\tiny{EB}}}\to-\boldsymbol{\chi}_{\text{\tiny{EB}}}\)~\footnote{To prove this take (\ref{field-G}) and write it as \(\hat{\boldsymbol{L}}_{\boldsymbol{x}}\boldsymbol{\cdot}\boldsymbol{G}(k,\boldsymbol{x}_{\parallel},\boldsymbol{x}_{\parallel}^{\prime},\Omega)=\boldsymbol{\mathbb{1}}_{3}\delta^{(2)}(\boldsymbol{x}_{\parallel}-\boldsymbol{x}_{\parallel}^{\prime})\), where \(\hat{\boldsymbol{L}}_{\boldsymbol{x}}\) is an abbreviated notation for the linear operator on the left hand side of the differential equation that acts on \(\boldsymbol{x}\).  Form the dot product between this equation and \(\boldsymbol{G}(k,\boldsymbol{x}_{\parallel}^{\prime\prime},\boldsymbol{x}_{\parallel},\Omega)\) and integrate over \(\boldsymbol{x}_{\parallel}\), \(\int d^{2}\boldsymbol{x}_{\parallel}\boldsymbol{G}(k,\boldsymbol{x}_{\parallel}^{\prime\prime},\boldsymbol{x}_{\parallel},\Omega)\boldsymbol{\cdot}\hat{\boldsymbol{L}}_{\boldsymbol{x}}\boldsymbol{\cdot}\boldsymbol{G}(k,\boldsymbol{x}_{\parallel},\boldsymbol{x}_{\parallel}^{\prime},\Omega)=\boldsymbol{G}(k,\boldsymbol{x}_{\parallel}^{\prime\prime},\boldsymbol{x}_{\parallel}^{\prime},\Omega)\).  An integration by parts can then shift the operation of the \(\hat{\boldsymbol{L}}_{\boldsymbol{x}}\) onto \(\boldsymbol{G}(k,\boldsymbol{x}_{\parallel}^{\prime\prime},\boldsymbol{x}_{\parallel},\Omega)\), and we can thereby infer the differential equation satisfied by the second index of the transposed Green function.  The result is the same differential equation but with the velocity multiplied by \(-1\).})
	\begin{multline}
		\left[\boldsymbol{G}(k,\boldsymbol{x}_{\parallel}^{\prime},\boldsymbol{x}_{\parallel},\Omega)\boldsymbol{\times}\overleftarrow{\boldsymbol{\nabla}}^{\star}\boldsymbol{\cdot}\boldsymbol{\mu}^{-1}(\boldsymbol{x}_{\parallel},\Omega_{-})\right]\boldsymbol{\times}\overleftarrow{\boldsymbol{\nabla}}^{\star}-\Omega^{2}\boldsymbol{G}(k,\boldsymbol{x}_{\parallel}^{\prime},\boldsymbol{x}_{\parallel},\Omega)\boldsymbol{\cdot}\boldsymbol{\epsilon}(\boldsymbol{x}_{\parallel},\Omega_{-})\\
		-i\Omega[\boldsymbol{G}(k,\boldsymbol{x}_{\parallel}^{\prime},\boldsymbol{x}_{\parallel},\Omega)\boldsymbol{\times}\overleftarrow{\boldsymbol{\nabla}}^{\star}\boldsymbol{\cdot}\boldsymbol{\chi}_{\text{\tiny{BE}}}(\boldsymbol{x}_{\parallel},\Omega_{-})-(\boldsymbol{G}(k,\boldsymbol{x}_{\parallel}^{\prime},\boldsymbol{x}_{\parallel},\Omega)\boldsymbol{\cdot}\boldsymbol{\chi}_{\text{\tiny{EB}}}(\boldsymbol{x}_{\parallel},\Omega_{-}))\boldsymbol{\times}\overleftarrow{\boldsymbol{\nabla}}^{\star}]=\boldsymbol{\mathbb{1}}_{3}\delta^{(2)}(\boldsymbol{x}_{\parallel}-\boldsymbol{x}_{\parallel}^{\prime})\label{transpose-green}
	\end{multline}
	The transverse Green function which appears in the expansion of the vector potential operator is  related to (\ref{field-G}) by
	 \begin{equation}
	 	\boldsymbol{G}_{\perp}(k,\boldsymbol{x}_{\parallel},\boldsymbol{x}_{\parallel}^{\prime},\Omega)=\int d^{2}\boldsymbol{x}_{\parallel 1}\tilde{\boldsymbol{\delta}}_{\perp}(k,\boldsymbol{x}_{\parallel}-\boldsymbol{x}_{\parallel 1})\boldsymbol{\cdot}\boldsymbol{G}(k,\boldsymbol{x}_{\parallel 1},\boldsymbol{x}_{\parallel}^{\prime},\Omega)\label{trans-G}
	 \end{equation}
	 with,  \(\tilde{\boldsymbol{\delta}}_{\perp}(k,\boldsymbol{x}_{\parallel}-\boldsymbol{x}^{\prime}_{\parallel})=\int dx \boldsymbol{\delta}_{\perp}(\boldsymbol{x}-\boldsymbol{x}^{\prime})e^{-ik(x-x^{\prime})} \).
	The following result will be important in proving that the commutators take the correct form,
	\begin{multline}
		\int d^{2}\boldsymbol{x}_{\parallel1}\boldsymbol{G}(k,\boldsymbol{x}_{\parallel},\boldsymbol{x}_{\parallel1},\Omega+V k)\boldsymbol{\cdot}\sum_{\lambda}\hat{\boldsymbol{O}}_{\lambda}^{\dagger}(k,\boldsymbol{x}_{\parallel1},\Omega)\boldsymbol{\cdot}\hat{\boldsymbol{O}}_{\lambda}(k,\boldsymbol{x}_{\parallel1},\Omega)\boldsymbol{\cdot}\boldsymbol{G}^{\dagger}(k,\boldsymbol{x}_{\parallel}^{\prime},\boldsymbol{x}_{\parallel1},\Omega+V k)\\
		=\frac{\gamma\Omega}{\pi i}\left[\boldsymbol{G}(k,\boldsymbol{x}_{\parallel},\boldsymbol{x}_{\parallel}^{\prime},\Omega+V k)-\boldsymbol{G}^{\dagger}(k,\boldsymbol{x}_{\parallel}^{\prime},\boldsymbol{x}_{\parallel},\Omega+V k)\right]\label{operator-sum}
	\end{multline}
	This can be proved from taking (\ref{transpose-green}) multiplied on the right by \(\boldsymbol{G}^{\dagger}\), subtracting the adjoint of (\ref{transpose-green}) multiplied on the left by \(\boldsymbol{G}\), and integrating over \(\boldsymbol{x}_{\parallel}\).  This can then be seen to equal (\ref{operator-sum}) after and application of (\ref{shorthand-operators}) and (\ref{susdef}).  Throughout these manipulations we must take care to remember that the symbol \(\boldsymbol{\nabla}\) contains a component \(i k\hat{\boldsymbol{x}}\) that reverses sign relative to \(\boldsymbol{\nabla}_{\parallel}\) when integrating by parts.  Explicitly if we take two vector fields, \(\boldsymbol{V}(k,\boldsymbol{x}_{\parallel})\) \& \(\boldsymbol{W}(k,\boldsymbol{x}_{\parallel})\) then, neglecting boundary terms, \(\int d^{2}\boldsymbol{x}_{\parallel} \boldsymbol{V}\boldsymbol{\cdot}\boldsymbol{\nabla}\boldsymbol{\times}\boldsymbol{W}=\int d^{2}\boldsymbol{x}_{\parallel} \boldsymbol{W}\boldsymbol{\cdot}\boldsymbol{\nabla}^{\star}\boldsymbol{\times}\boldsymbol{V}\equiv\int d^{2}\boldsymbol{x}_{\parallel} \boldsymbol{V}\boldsymbol{\times}\overleftarrow{\boldsymbol{\nabla}}^{\star}\boldsymbol{\cdot}\boldsymbol{W}\).
	\par
	The remaining expansion coefficients are as follows: the canonical electromagnetic field momentum expansion coefficient is given by
	 \begin{multline}
	 \boldsymbol{f}^{\lambda}_{\Pi_{A}}(k,\boldsymbol{x}_\parallel,\boldsymbol{x}_{\parallel 1},\Omega)=-\delta^{(2)}(\boldsymbol{x}_{\parallel}-\boldsymbol{x}_{\parallel1})\boldsymbol{\alpha}_{\text{\tiny{E$\lambda$}}}(\boldsymbol{x}_{\parallel1},\gamma\Omega)\\
		-\bigg[i(\Omega+Vk)\boldsymbol{\epsilon}(\boldsymbol{x}_{\parallel},\gamma\Omega)+\boldsymbol{\chi}_{\text{\tiny{EB}}}(\boldsymbol{x}_{\parallel},\gamma\Omega)\boldsymbol{\cdot}\boldsymbol{\nabla}\boldsymbol{\times}\bigg]\boldsymbol{\cdot}\boldsymbol{G}(k,\boldsymbol{x}_{\parallel},\boldsymbol{x}_{\parallel1},\Omega+Vk)\boldsymbol{\cdot}\hat{\boldsymbol{O}}^{\dagger}_{\lambda}(k,\boldsymbol{x}_{\parallel1},\Omega)\label{Pi-A-expansion}
	 \end{multline}
	the expansion coefficients for the oscillator field operators are
	 \begin{multline}
	 	\boldsymbol{f}^{\lambda}_{X_{\omega}}(k,\boldsymbol{x}_\parallel,\boldsymbol{x}_{\parallel 1},\Omega)=\boldsymbol{\mathbb{1}}_{3}\delta_{\text{\tiny{$\lambda$E}}}\delta(\gamma\Omega-\omega)\delta^{(2)}(\boldsymbol{x}_{\parallel}-\boldsymbol{x}_{\parallel1})\\
		+\frac{\left[i(\Omega+Vk)\boldsymbol{\alpha}^{T}_{\text{\tiny{EE}}}(\boldsymbol{x}_{\parallel},\omega)+\boldsymbol{\alpha}^{T}_{\text{\tiny{BE}}}(\boldsymbol{x}_{\parallel},\omega)\boldsymbol{\cdot}\boldsymbol{\nabla}\boldsymbol{\times}\right]\boldsymbol{\cdot}\boldsymbol{G}(k,\boldsymbol{x}_{\parallel},\boldsymbol{x}_{\parallel1},\Omega+Vk)\boldsymbol{\cdot}\hat{\boldsymbol{O}}^{\dagger}_{\lambda}(k,\boldsymbol{x}_{\parallel1},\Omega)}{(\omega-\gamma\Omega-i\eta)(\omega+\gamma\Omega+i\eta)}.\label{X-expansion}
	 \end{multline}
	 and the expansion coefficients for canonical momentum operator for the oscillator field are
	 \begin{equation}
	 	\boldsymbol{f}^{\lambda}_{\Pi_{X_{\omega}}}(k,\boldsymbol{x}_\parallel,\boldsymbol{x}_{\parallel 1},\Omega)=-i\gamma^{2}\Omega\boldsymbol{f}^{\lambda}_{X_{\omega}}(k,\boldsymbol{x}_\parallel,\boldsymbol{x}_{\parallel 1},\Omega).\label{Pi-X-expansion}
	 \end{equation}
	 We now insert (\ref{A-expansion}), (\ref{Pi-A-expansion}), (\ref{X-expansion}) \& (\ref{Pi-X-expansion}) into (\ref{EM-integral}--\ref{X-integral}).  In the case of the canonical commutation relation for the electromagnetic field (\ref{EM-integral}) this gives,
	\begin{multline}
		\int_{-\infty}^{\infty}\frac{dk}{2\pi}e^{ik(x-x^{\prime})}\int_{-\infty}^{\infty}\frac{d\Omega}{2\pi}\bigg\{\boldsymbol{G}_{\perp}(k,\boldsymbol{x}_{\parallel},\boldsymbol{x}_{\parallel}^{\prime},\Omega+Vk)\boldsymbol{\cdot}\left[-i(\Omega+Vk)\boldsymbol{\epsilon}(\boldsymbol{x}_{\parallel}^{\prime},\gamma\Omega)+\boldsymbol{\times}\stackrel{\leftarrow}{\boldsymbol{\nabla}}^{\star\prime}\boldsymbol{\cdot}\boldsymbol{\chi}_{\text{\tiny{BE}}}(\boldsymbol{x}_{\parallel}^{\prime},\gamma\Omega)\right]\\
		-\boldsymbol{G}^{\dagger}_{\perp}(k,\boldsymbol{x}_{\parallel}^{\prime},\boldsymbol{x}_{\parallel},\Omega+Vk)\boldsymbol{\cdot}\left[-i(\Omega+Vk)\boldsymbol{\epsilon}^{\star}(\boldsymbol{x}_{\parallel}^{\prime},\gamma\Omega)+\boldsymbol{\times}\stackrel{\leftarrow}{\boldsymbol{\nabla}}^{\prime}\boldsymbol{\cdot}\boldsymbol{\chi}_{\text{\tiny{BE}}}^{\star}(\boldsymbol{x}_{\parallel}^{\prime},\gamma\Omega)\right]\bigg\}\stackrel{\text{?}}{=}\boldsymbol{\delta}_{\perp}(\boldsymbol{x}-\boldsymbol{x}^{\prime})\label{em-comm-res}
	\end{multline}
	where we have applied (\ref{susdef}) and (\ref{operator-sum}).  The adjoint of the Green function, \(\boldsymbol{G}^{\dagger}\) in (\ref{em-comm-res})  is transverse with respect to the left hand index.
	\par
	The Green functions and the susceptibilities are both analytic functions of frequency in the upper half plane, with the complex conjugates of these quantities being analytic in the lower half plane.  Therefore we can split the integrand of (\ref{em-comm-res}) into two pieces---one analytic in the upper half plane, and one in the lower half.  We can then deform the integration contour over frequency from being along the real line, to following a semicircular path from \(-\infty\) to \(\infty\), in either the clockwise (\(\mathcal{C}^{+}\)) or anticlockwise (\(\mathcal{C}^{-}\)) sense, depending on where the integrand is analytic.  Along \(\mathcal{C}^{\pm}\) we have, \(|\Omega|\to\infty\), where the Green function can be taken to satisfy the free space equation and is therefore equal to
	\begin{equation}
		\boldsymbol{G}(k,\boldsymbol{x}_{\parallel},\boldsymbol{x}_{\parallel}^{\prime},\Omega)\to-\mu_{0}\int\frac{d^{2}\boldsymbol{K}}{(2\pi)^{2}}\left(\frac{\frac{\Omega^{2}}{c^{2}}\boldsymbol{\mathbb{1}}_{3}-\boldsymbol{k}\boldsymbol{\otimes}\boldsymbol{k}}{\frac{\Omega^{2}}{c^{2}}\left(\frac{\Omega^{2}}{c^{2}}-\boldsymbol{k}^{2}\right)}\right)e^{i\boldsymbol{K}\boldsymbol{\cdot}(\boldsymbol{x}_{\parallel}-\boldsymbol{x}_{\parallel}^{\prime})}\label{green-limit}
	\end{equation}
	where \(\boldsymbol{k}=k\hat{\boldsymbol{x}}+\boldsymbol{K}\).  As \(\boldsymbol{k}\) is a real vector, the denominator in (\ref{green-limit}) is non--zero over the entire contour.  Therefore, \(\boldsymbol{G}(k,\boldsymbol{x}_{\parallel},\boldsymbol{x}_{\parallel}^{\prime},\Omega)\to-\frac{1}{\epsilon_{0}\Omega^{2}}\boldsymbol{\mathbb{1}}_{3}\delta^{(2)}(\boldsymbol{x}_{\parallel}-\boldsymbol{x}_{\parallel}^{\prime})\), and (\ref{em-comm-res}) simplifies to,
	\begin{equation}
		\int_{-\infty}^{\infty}\frac{dk}{2\pi}e^{ik(x-x^{\prime})}\tilde{\boldsymbol{\delta}}_{\perp}(k,\boldsymbol{x}_{\parallel}-\boldsymbol{x}_{\parallel}^{\prime})\left[-\int_{\mathcal{C}^{+}}\frac{d\Omega}{2\pi\Omega}+\int_{\mathcal{C}^{-}}\frac{d\Omega}{2\pi\Omega}\right]\stackrel{\text{?}}{=}i\boldsymbol{\delta}_{\perp}(\boldsymbol{x}-\boldsymbol{x}^{\prime})\label{em-comm-res-2}
	\end{equation}
	writing  \(\Omega=|\Omega|e^{i\theta}\), and performing the two contour integrals in the square brackets gives a factor of \(i\), and the equality is fulfilled.  Therefore the expansion of the electromagnetic field operators in the form (\ref{op-expansion}) with the coefficients (\ref{A-expansion}) and (\ref{Pi-A-expansion}) is consistent with the canonical commutation relations.
	\par
	For the oscillator fields, inserting (\ref{X-expansion}) and (\ref{Pi-X-expansion}) into (\ref{X-integral}), we find that after applying (\ref{operator-sum}), all terms cancel in the product [\(\boldsymbol{f}^{\lambda}_{X_{\omega}}(k,\boldsymbol{x}_\parallel,\boldsymbol{x}_{\parallel 1},\Omega)\boldsymbol{\cdot}\boldsymbol{f}^{\lambda\,\dagger}_{\Pi_{X_{\omega^{\prime}}}}(k,\boldsymbol{x}_{\parallel}^{\prime},\boldsymbol{x}_{\parallel1},\Omega)-\text{c.c.}]\) except for the first involving products of delta functions.  We are then left with
	\begin{multline}
		\sum_{\lambda}\int\frac{dk}{2\pi}\int d^{2}\boldsymbol{x}_{\parallel1}\int_{0}^{\infty}\frac{d\Omega}{2\gamma\Omega}\left[i\gamma^{2}\Omega e^{ik(x-x^{\prime})}\boldsymbol{\mathbb{1}}_{3}\delta_{\text{\tiny{$\lambda$E}}}\delta_{\text{\tiny{$\lambda$E}}}\delta(\gamma\Omega-\omega)\delta(\gamma\Omega-\omega^{\prime})\delta^{(2)}(\boldsymbol{x}_{\parallel}-\boldsymbol{x}_{\parallel1})\delta^{(2)}(\boldsymbol{x}_{\parallel}^{\prime}-\boldsymbol{x}_{\parallel1})-\text{c.c.}\right]\\
		\stackrel{?}{=}i\boldsymbol{\mathbb{1}}_{3}\delta(\omega-\omega^{\prime})\delta^{(3)}(\boldsymbol{x}-\boldsymbol{x}^{\prime})\label{final-osc-comm}
	\end{multline}
	Performing the integrals and summation on the left of (\ref{final-osc-comm}) shows that the equality is fulfilled, and this completes our demonstration that (\ref{quant-class}) is a canonical transformation.
%
%
	\section{Including the centre of mass as a dynamical variable\label{appendix-C}}
	\par
	A natural reaction to a Hamiltonian for a quantum field that includes negative energy excitations, such as (\ref{quantum-hamiltonian}), is that it is physically unacceptable.  If the Hamiltonian included all of the possible dynamical degrees of freedom as canonical variables then this would probably be true.  However, in our case the centre of mass velocity of the dielectric medium is assumed to be fixed.  We might therefore interpret the negative energy as an accounting device for the external energy input required to keep the motion constant.  To show that this interpretation is reasonable, we now demonstrate that including the centre of mass as a dynamical variable removes the terms that cause (\ref{quantum-hamiltonian}) to be unbounded from below.
	\par
	We start from (\ref{lagrangian-density}), adding in an extra term to represent the kinetic energy of the centre of mass.  The centre of mass is only a meaningful concept to first order in \(\boldsymbol{V}/c\), and we therefore assume a motion of the medium such that we can set \(\gamma\sim1\).  For arbitrary motion we should use the centre of energy, and the treatment becomes much less straightforward.  To this order the total Lagrangian is,
	\[
		L=\frac{1}{2}M\boldsymbol{V}^{2}+\int d^{3}\boldsymbol{x}\left[\mathcal{L}_{\text{\tiny{F}}}+\mathcal{L}_{\text{\tiny{R}}}+\mathcal{L}_{\text{\tiny{INT}}}\right]
	\]
	For these purposes, we assume that \(\boldsymbol{V}=V\hat{\boldsymbol{x}}\) as in the main text, and construct the Hamiltonian as in section~\ref{classical-hamiltonian}.  The canonical momentum associated with the centre of mass is,
	\[
		\boldsymbol{P}=\frac{\partial L}{\partial\boldsymbol{V}}=M\boldsymbol{V}+\int d^{3}\boldsymbol{x}\int_{0}^{\infty}d\omega\left[\boldsymbol{\nabla}\boldsymbol{\otimes}\boldsymbol{X}_{\omega}\boldsymbol{\cdot}\boldsymbol{\Pi}_{X_{\omega}}-\alpha_{\text{\tiny{EE}}}(\boldsymbol{x}_{\parallel},\omega)\boldsymbol{X}_{\omega}\boldsymbol{\times}\boldsymbol{B}+\alpha_{\text{\tiny{BB}}}(\boldsymbol{x}_{\parallel},\omega)\boldsymbol{Y}_{\omega}\boldsymbol{\times}\boldsymbol{E}/c^{2}\right]	
	\]
	and therefore the Hamiltonian is,

	\begin{equation}
		H=\frac{1}{2}M\boldsymbol{V}^{2}+\int d^{3}\boldsymbol{x}\left\{\frac{\epsilon_{0}}{2}\left(\boldsymbol{E}^{2}+c^{2}\boldsymbol{B}^{2}\right)+\frac{1}{2}\int_{0}^{\infty}d\omega\left[\boldsymbol{\Pi}_{X_{\omega}}^{2}+\boldsymbol{\Pi}_{Y_{\omega}}^{2}+\omega^{2}\left(\boldsymbol{X}_{\omega}^{2}+\boldsymbol{Y}_{\omega}^{2}\right)-\alpha_{\text{\tiny{BB}}}(\omega)\boldsymbol{Y}_{\omega}\boldsymbol{\cdot}\left(\boldsymbol{B}+\boldsymbol{V}\boldsymbol{\times}\boldsymbol{E}/c^{2}\right)\right]\right\}\label{vel-ham}
	\end{equation}
	The term, \(\boldsymbol{V}\boldsymbol{\cdot}(\boldsymbol{\nabla}\boldsymbol{\otimes}\boldsymbol{X}_{\omega})\boldsymbol{\cdot}\boldsymbol{\Pi}_{X_{\omega}}\) present in (\ref{oscillator-hamiltonian}) has been removed from (\ref{vel-ham}) through the inclusion of the centre of mass dynamics.  As explained in the main text, it is this term that is responsible for the lack of a lower bound to the Hamiltonian.  It is not possible for (\ref{vel-ham}) to be decreased to an arbitrary negative value.
%
%
	\section{The transition rate of a detector interacting with a moving medium\label{appendix-D}}
	\par
	To find the transition rate, we first find the probability amplitude associated with the atom making a transition into an excited state, given that the field is also in some given state.  In (\ref{expansion-change}) it was found that the rate of change of this amplitude is given by
	\[
		\dot{\zeta}_{\lambda,\sigma,l,m,n,p}(t)=\frac{1}{\hbar}\langle1_{\lambda,\sigma,l,m,n,p}|\boldsymbol{\kappa}\boldsymbol{\cdot}\hat{\boldsymbol{E}}(\boldsymbol{x}_{0},t)e^{i\omega t}|0\rangle
	\]
	where the matrix element associated with the atomic transition has been calculated.  To calculate the matrix element associated with the combined system of the electromagnetic field and the dielectric medium we use the expansion of the electric field in the form given in (\ref{op-expansion}).  We then expand the \(\hat{\boldsymbol{C}}_{\lambda}\) operators in terms of the \(\hat{c}_{\sigma\lambda}\) operators using (\ref{completeness-relation}),
	\[
		\hat{\boldsymbol{C}}_{\lambda}(k,\boldsymbol{x}_{\parallel}^{\prime},\gamma\Omega)=\sum_{\sigma}\boldsymbol{e}_{\sigma}\sum_{l,m,n,p}\phi^{\star}_{l,m,n,p}(k,\boldsymbol{x}_{\parallel}^{\prime},\gamma\Omega)\hat{c}_{\lambda,\sigma}(l,m,n,p)
	\]
	The orthonormality of the number states given in (\ref{number-states}) is then applied to give
	\[
		\dot{\zeta}_{\lambda,\sigma,l,m,n,p}(t)=\frac{1}{\hbar}\boldsymbol{\kappa}\boldsymbol{\cdot}\int\frac{dk}{2\pi}\int d^{2}\boldsymbol{x}_{\parallel}^{\prime}\int_{0}^{\infty}d\Omega\sqrt{\frac{\hbar}{2\Omega}}{\boldsymbol{f}_{\text{\tiny{E}}}^{\lambda}}^{\star}(k,\boldsymbol{x}_{\parallel0},\boldsymbol{x}_{\parallel}^{\prime},\Omega)\boldsymbol{\cdot}\boldsymbol{e}_{\sigma}\phi_{l,m,n,p}(k,\boldsymbol{x}_{\parallel}^{\prime},\gamma\Omega)e^{i[-kx_{0}+(\Omega+V k+\omega)t]}
	\]
	Integrating this over a time interval \([0,T]\), squaring the resulting expression and summing over all possible final states then gives
	\begin{equation}
		\sum_{\lambda,\sigma,l,m,n,p}|\zeta_{\lambda,\sigma,l,m,n,p}(T)|^{2}=\int\frac{dk}{2\pi}\int d^{2}\boldsymbol{x}_{\parallel}^{\prime}\int_{0}^{\infty}d\Omega\sum_{\lambda}\left|\boldsymbol{\kappa}\boldsymbol{\cdot}\boldsymbol{f}_{\text{\tiny{E}}}^{\lambda}(k,\boldsymbol{x}_{\parallel0},\boldsymbol{x}_{\parallel}^{\prime},\Omega)\right|^{2}\frac{2\sin^{2}[(\Omega+V k+\omega)T/2]}{\gamma\hbar\Omega(\Omega+V k+\omega)^{2}}\label{transprob}
	\end{equation}
	where we assume that the value of the expansion coefficient is equal to zero at \(t=0\), and we have applied the completeness relation (\ref{completeness-relation}).  Dividing (\ref{transprob}) by the time interval, \(T\) gives the transition rate, \(\Gamma_{0\to1}\).  As \(T\to\infty\), we can apply the identity
	\begin{equation}
		\lim_{T\to\infty} \frac{4\sin^{2}[(\Omega+V k+\omega)T/2]}{(\Omega+V k+\omega)^{2}T}=2\pi\delta[(\Omega+V k+\omega)]\label{delta-identity}
	\end{equation}
	and the transition rate simplifies to
		\begin{equation}
		\Gamma_{0\to1}=-\frac{1}{2\hbar}\int_{-\infty}^{-\omega/V}dk\int d^{2}\boldsymbol{x}_{\parallel}\sum_{\lambda}\frac{1}{\omega_{+}}\left|\boldsymbol{\kappa}\boldsymbol{\cdot}\boldsymbol{f}_{\text{\tiny{E}}}^{\lambda}(k,\boldsymbol{x}_{\parallel0},\boldsymbol{x}_{\parallel},-\omega-V k)\right|^{2}
	\end{equation}
	Inserting (\ref{E-expansion}) and applying (\ref{operator-sum}) then gives the transition rate in terms of the Green function,
	\begin{equation}
		\Gamma_{0\to1}=-\frac{2\omega^{2}}{\hbar}\int_{\omega/V}^{\infty}\frac{dk}{2\pi}\boldsymbol{\kappa}\boldsymbol{\cdot}\frac{\boldsymbol{G}(k,\boldsymbol{x}_{\parallel0},\boldsymbol{x}_{\parallel0},\omega)-\boldsymbol{G}^{\dagger}(k,\boldsymbol{x}_{\parallel0},\boldsymbol{x}_{\parallel0},\omega)}{2i}\boldsymbol{\cdot}\boldsymbol{\kappa}\label{transeq2}
	\end{equation}
	When a medium is reciprocal then it satisfies (\ref{reciprocity}), and \(\boldsymbol{G}^{T}(k,\boldsymbol{x}_{\parallel0},\boldsymbol{x}_{\parallel0},\omega)=\boldsymbol{G}(k,\boldsymbol{x}_{\parallel0},\boldsymbol{x}_{\parallel0},\omega)\).  As stated in appendix \ref{app-B}, for moving media (\ref{reciprocity}) is not satisfied, and we might expect this is affect \(\Gamma_{0\to1}\).  Yet apparently the non--reciprocity has no effect on the transition rate of interest here (\ref{transeq2}), for we contract the indices of the Green tensor with the symmetric tensor \(\boldsymbol{\kappa}\boldsymbol{\otimes}\boldsymbol{\kappa}\), which gives
		\begin{equation}
		\Gamma_{0\to1}=-\frac{2\omega^{2}}{\hbar}\int_{\omega/V}^{\infty}\frac{dk}{2\pi}\boldsymbol{\kappa}\boldsymbol{\cdot}\text{Im}[\boldsymbol{G}(k,\boldsymbol{x}_{\parallel0},\boldsymbol{x}_{\parallel0},\omega)]\boldsymbol{\cdot}\boldsymbol{\kappa}=\frac{2\omega^{2}}{\hbar}\int_{\omega/V}^{\infty}\frac{dk}{2\pi}\boldsymbol{\kappa}\boldsymbol{\cdot}\text{Im}[\boldsymbol{G}(-k,\boldsymbol{x}_{\parallel0},\boldsymbol{x}_{\parallel0},-\omega)]\boldsymbol{\cdot}\boldsymbol{\kappa}\label{finaltranseq}
	\end{equation}
	which is the expression given in the text (\ref{transition-rate}).
	\end{widetext}
	\bibliography{refs}

\begin{thebibliography}{10}

\bibitem{fulling__1976}
S.~A. Fulling and P.~C.~W. Davies.
\newblock {\em Proc. Roy Soc. A}, 348:393, 1976.

\bibitem{hawking__1975}
S.~W. Hawking.
\newblock {\em Commun. Math. Phys.}, 43:199, 1975.

\bibitem{schutzhold__2005}
R.~Sch\"utzhold and W.~G. Unruh.
\newblock {\em Phys. Rev. Lett}, 95:031301, 2005.

\bibitem{philbin__2008}
T.~G. Philbin, C.~Kuklewicz, S.~Robertson, S.~Hill, F.~K\"onig, and
  U.~Leonhardt.
\newblock {\em Science}, 319:1367, 2008.

\bibitem{belgiorno__2010}
F.~Belgiorno, S.~L. Cacciatori, M.~Clerici, V.~Gorini, G.~Ortenzi, L.~Rizzi,
  E.~Rubino, V.~G. Sala, and D.~Faccio.
\newblock {\em Phys. Rev. Lett.}, 105:203901, 2010.

\bibitem{pendry__1997}
J.~B. Pendry.
\newblock {\em J. Phys. Cond. Mat.}, 9:1703, 1997.

\bibitem{pendry__2010}
J.~B. Pendry.
\newblock {\em New J. Phys.}, 12:068002, 2010.

\bibitem{leonhardt__2010}
U.~Leonhardt.
\newblock {\em New J. Phys.}, 12:068001, 2010.

\bibitem{barton2010a}
G.~Barton.
\newblock {\em New J. Phys.}, 12:113044, 2010.

\bibitem{barton2010b}
G.~Barton.
\newblock {\em New J. Phys.}, 12:113045, 2010.

\bibitem{volokitin2011}
A.~I. Volokitin and B.~N.~J. Persson.
\newblock {\em Phys. Rev. Lett.}, 106:094502, 2011.

\bibitem{schutzhold__2011}
R.~Sch\"utzhold and W.~G. Unruh.
\newblock {\em Phys. Rev. Lett.}, 107:149401, 2011.

\bibitem{belgiorno__2011}
F.~Belgiorno, S.~L. Cacciatori, M.~Clerici, V.~Gorini, G.~Ortenzi, L.~Rizzi,
  E.~Rubino, V.~G. Sala, and D.~Faccio.
\newblock {\em Phys. Rev. Lett.}, 107:149402, 2011.

\bibitem{jauch__1948}
J.~M. Jauch and K.~M. Watson.
\newblock {\em Phys. Rev.}, 74:950, 1948.

\bibitem{volume5}
L.~D. Landau and E.~M. Lifshitz.
\newblock {\em Statistical Physics - Part 1}.
\newblock {Butterworth-Heinemann}, Oxford, 2005.

\bibitem{huttner__1992}
B.~Huttner and S.~M. Barnett.
\newblock {\em Phys. Rev. A}, 46:4306, 1992.

\bibitem{philbin__2010}
T.~G. Philbin.
\newblock {\em New J. Phys.}, 12:123008, 2010.

\bibitem{horsley__2011-1}
S.~A.~R. Horsley.
\newblock {\em Phys. Rev. A}, 84:063822, 2011.

\bibitem{matloob2005b}
R.~Matloob.
\newblock {\em Phys. Rev. A}, 71:062105, 2005.

\bibitem{berestetskii_quantum_2004}
V.~B. Berestetskii, E.~M. Lifshitz, and L.~P. Pitaevskii.
\newblock {\em Quantum Electrodynamics}.
\newblock {Butterworth-Heinemann}, Oxford, 2004.

\bibitem{craig_molecular_1998}
D.~P. Craig and T.~Thirunamachandran.
\newblock {\em Molecular Quantum Electrodynamics}.
\newblock Dover, New York, 1998.

\bibitem{loudon1983}
R.~Loudon.
\newblock {\em The Quantum Theory of Light}.
\newblock Oxford University Press, 1983.

\bibitem{ackerhalt1984}
J.~R. Ackerhalt and P.~W. Milonni.
\newblock {\em J. Opt. Soc. Am. B}, 1:116, 1984.

\bibitem{barnett1992}
S.~M. Barnett, B.~Huttner, and R.~Loudon.
\newblock {\em Phys. Rev. Lett.}, 68:3698, 1992.

\bibitem{barnett1996}
S.~M. Barnett, B.~Huttner, R.~Loudon, and R.~Matloob.
\newblock {\em J. Phys. B: At. Mol. Opt. Phys.}, 29:3763, 1996.

\bibitem{dung2000}
H.~T. Dung, L.~Knoll, and D-G Welsch.
\newblock {\em Phys. Rev. A}, 62:053804, 2000.

\bibitem{blow__1990}
K.~J. Blow, R.~Loudon, S.~J.~D. Phoenix, and T.~J. Shepherd.
\newblock {\em Phys. Rev. A}, 42(7):4102--4114, 1990.

\bibitem{landau_quantum_2003}
L.~D. Landau and E.~M. Lifshitz.
\newblock {\em Quantum Mechanics}.
\newblock {Butterworth-Heinemann}, Oxford, 2003.

\bibitem{cohen2011}
A.~G. Cohen and S.~L. Glashow.
\newblock {\em Phys. Rev. Lett.}, 107:181803, 2011.

\bibitem{scheel2009}
S.~Scheel and S.~Y. Buhmann.
\newblock {\em Phys. Rev. A}, 80:042902, 2009.

\bibitem{horsley2011b}
S.~A.~R. Horsley, M.~Artoni, and G.~C. La~Rocca.
\newblock {\em arXiv:1111.4352v1}, 2011.

\bibitem{amooshahi__2009}
M.~Amooshahi.
\newblock {\em Eur, Phys. J. D}, 54:115, 2009.

\end{thebibliography}
\end{document}